\documentclass[12pt,a4paper]{article}
\pdfoutput=1
\usepackage{comment}
\usepackage{amsmath}
\usepackage{graphicx}
\usepackage{caption}
\usepackage{subcaption}
\usepackage{jheppub}
\usepackage{enumitem}
\usepackage{framed,float}
\usepackage{booktabs}
\usepackage{subcaption}
\usepackage{tikz}
\usetikzlibrary{calc,arrows,decorations.markings}
\usepackage{mathrsfs}
\usepackage[utf8]{inputenc}
\usepackage{ytableau}
\usepackage{graphicx}

\newcommand{\R}{\mathbb{R}}
\newcommand{\Z}{\mathbb{Z}}

\usepackage{color}

\usepackage{siunitx,amsmath}

\newcommand{\itemEq}[1]{%
	\begingroup%
	\setlength{\abovedisplayskip}{0pt}%
	\setlength{\belowdisplayskip}{0pt}%
	\parbox[c]{\linewidth}{\begin{flalign}#1&&\end{flalign}}%
	\endgroup}

\title{On The Finiteness of 6d Supergravity Landscape }

\author[a]{Houri-Christina Tarazi,\,}
\author[a]{Cumrun Vafa\,}

\affiliation[a]{Department of Physics, Harvard University, Cambridge, MA 02138, USA}

\abstract
{
We consider supergravity theories with 8 supercharges in $d=6$.  We show that all the proposed anomaly free theories with unbounded number of massless modes are restricted to a finite subset and thus argue   that there is an upper bound on the number of massless modes, consistent with the String Lamppost Principle.
}

\begin{document}
\begin{flushright}
\tt 

\end{flushright}

\maketitle

\section{Introduction}\label{sec:intro}
\ytableausetup{boxsize=2em}\
It has been well established that the existence of a UV complete quantum gravity theory puts strict constraints on the set of quantum field theory that can appear in the low-energy. Such consistency conditions are said to divide the consistent quantum field theories to those in the Landscape (consistent when coupled to gravity) and those not consistent which are said to belong to the Swampland.  The Swampland program  \cite{Vafa:2005ui} attempts to identify such conditions independently of the particular UV theory.  Therefore, it would be interesting to evaluate the validity of the String Lamppost Principle, i.e. the idea that all consistent quantum gravitational theories are part of the string theory landscape.  An interesting first question one could ask is if the string Swampland condition that there is an upper bound on the number of  massless modes \cite{Vafa:2005ui} is a consequence of consistency of arbitrary consistent theories of quantum gravity. This conjecture is motivated from the fact that only a finite number of Calabi-Yau compactifications \cite{Yau:1991} is believed to exist for a specific dimension and amount of supersymmetry and hence leading to a finite set of supersymmetric quantum gravitational theories with a bounded set of massless modes.
 
To try and answer this questions we can firstly look at the simplest example of such theories. Our first stop are the  maximal supersymmetric theories with 32 supercharges whose massless modes are determined by supersymmetry considerations and lead to a finite massless spectrum.  In fact all these theories are realized in string theory and hence provide an example of the String Lamppost Principle(SLP). The next stop are supersymmetric theories with 16 supercharges {\cite{Adams:2010zy,Green:1984sg, Kim:2019aa}}, which were recently shown \cite{Kim_2020d} to enjoy an upper bound on their rank $r\leq 26-d$, a bound also first suggested by string theory.  This therefore completes the finiteness arguments for the number of massless modes in supergravity theories with 16 supercharges.

To further check this finiteness hypothesis, it is only natural to move to our next stop:  theories with 8 supercharges. Such an amount of supersymmetry firstly appears in 6 dimensions which also constitutes the simplest dimension to study also because of the constraints imposed by chiral anomaly  cancellations. Over the past decades a lot of effort on analyzing the Landscape of such theories \cite{Kumar_2010,Taylor_2019, Kumar_2009, Kumar_2011, Morrison_2012, Morrison_2012b, Kumar_2010b, raghuram2020automatic, Kim:2019aa,  Lee_2019, Park_2012} has led to a better understanding of the possible consistent theories, although not yet complete.  

 As mentioned above it is crucial to wonder whether at least the boundedness of the massless modes is true in such theories.  In this paper we will first summarize the classes of potential theories for which this is known to be true and then extend it to conclude that at least for all the proposed classes the number of massless modes are bounded. 
 
A second step to understanding the SLP is whether all theories in the Landscape can also be constructed in string theory. In particular in the case of 32 supercharges one can show that the only consistent theories are unique (except 10d, IIB and IIA) in each dimension and can be constructed in string theory. In theories with 16 supercharges progress has been made to show that the ranks of gauge groups that could consistently appear \cite{Cveti__2020,Dierigl_2021} indicate a tendency to match those coming from string theory. In particular, in 9 dimensions the only ranks that appear are $1,9$ and $17$ while for 8 dimensions the only ranks that appear are $2,10$ and $18$. In \cite{AlvarezGaume:1983ig} it is shown that indeed the rank should be odd in 9d and even in 8d otherwise the theory would suffer from a global gravitational anomaly. A more refined work was done in  \cite{Montero_2021} to explain this exact pattern of numbers appearing using the cobordism conjecture. The next question one could ask is what specific gauge groups could appear and whether they match those coming from string theory. In particular, in \cite{Garc_a_Etxebarria_2017} it is shown  that $f_4,so(2N+1)$ gauge groups which have no string theory constructions are in fact anomalous in 8d. More recently, \cite{hamada20218d} shows that the gauge group $g_2$ which also has no string theory construction, in fact belongs to the Swampland by  considering 3-brane probes. A further analysis of possible gauge groups in 8d  has been conducted in \cite{Cveti__2020,font2021exploring,Font_2020}.

  Moreover, It would be natural to  ask whether this is also the case for theories with 8 supercharges.  We thus need to understand more conditions to further constrain the Landscape.  In this work we will try to add another consistency condition that the quantum field theories need to satisfy coming from the completeness of spectrum hypothesis.  In particular we identify constraints coming from unitarity constraints on the 2d worldsheet of BPS string on the type of bulk matter representations that can appear. This is interesting because apart from anomaly arguments, there are no previous analyses that constrain the type of matter that can appear.  Of course in the string landscape one sees strong restrictions on possible matter content coming from F-theory \cite{Klevers_2017,Kumar_2010b, Morrison_2012f}.

The organization of this paper is as follows: In section 2 we review known consistency conditions for 6d ${\cal N}=1$ supergravity theories. In section 3 
we review known examples and construct new classes of potential anomaly free 6d theories with 8 supercharges, with an eye towards families which are not bounded in the  number of massless modes.  We then move on to show that all such theories are restricted to a finite range based on unitarity of BPS string probes.  In section 4 we discuss novel restrictions coming from unitarity of BPS string probes on the matter representations and use that to rule out one theory which did not have any known string theory construction.  In section 5 we discuss some directions for future research.  Lastly, some technical aspects are presented in the Appendices.

 \section{Review of 6d $\mathcal{N}=1$ Supergravity }\label{sec:1}
  \ytableausetup{boxsize=0.7 em,aligntableaux = center}
 In this section we review various  features of  6d $(1,0)$ supergravity theories. In addition, we provide a review of the  conditions that have been conjectured  to be necessary for the consistency of these theories. The set of all such conditions provide Swampland constraints that severely limit the possible quantum field theory that could arise in consistent quantum gravity theories.

  	\textbf{Anomaly Cancellation consideration :}
 A six-dimensional supergravity with 8 supercharges consists of four types of massless supermultiplets: a gravity multiplet, vector multiplets, tensor multiplets and hypermultiplets.  The chiral fields of those multiplets contribute to the anomalies produced in such a  theory characterized by an 8-form anomaly polynomial $I_8$. Such anomalies can be cancelled by the Green-Schwarz-Sagnotti mechanism \cite{Sagnotti:1992qw} if the anomaly polynomial $I_8$ factorizes as 
 \begin{eqnarray}
 I_8(R,F)={1\over 2 }\Omega_{\alpha\beta}X^\alpha_4 X^\beta_4, \ \ X_4^\alpha={1\over 2 }a^\alpha trR^2 +\sum_ib_i^\alpha {2\over \lambda_i }trF_i^2
 \end{eqnarray}
 where $a^\alpha, b_i^\alpha$ are vectors in $\R^{1,T}$, $\Omega_{\alpha \beta }$ is the metric on this space and  $ \lambda_i $ are normalization factors of the gauge groups $G_i$.
 The anomaly factorization conditions for gravitational, gauge and mixed anomalies are summarized as follows:

 \begin{itemize}
 	\item $\itemEq{\label{eqn:R4} R^4:  \ \ H-V=273-29T}$
 	\item$ \itemEq{\label{eqn:F4}F^4: \ \ 0=B^i_{Adj}-\sum n_R^i B^i_R}$
 	\item $\itemEq{\label{eqn:R22}(R^2)^2: \  a\cdot a=a^\alpha\Omega_{\alpha \beta }a^\beta  =9-T}$
 	\item $\itemEq{\label{eqn:F2R2} F^2R^2: \ a\cdot b_i=a^\alpha\Omega_{\alpha \beta }b_i^\beta   ={1\over 6 }\lambda_i (A^i_{Adj}-\sum_Rn_R^iA^i_R)} $
 	\item $\itemEq{\label{eqn:F22}(F^2)^2: \  b_i \cdot b_i =b_i^\alpha\Omega_{\alpha \beta }b_i^\beta  ={1\over 3}\lambda_i^2 (\sum_R n_R^iC^i_R-C^i_{Adj}) }$
 	\item $\itemEq{\label{eqn:F2F2}F^2_iF^2_j: \ b_i \cdot b_j=b_i^\alpha\Omega_{\alpha \beta }b_j^\beta  = \sum_{R,S}\lambda_i \lambda_jn_{RS}^{ij}A^i_RA^j_S  \ \ \ i\neq j }$
 	 \end{itemize}
 	where $H, V, T$  denote the number of hypermultiplets, vectors multiplets and tensor multiplets in the theory respectively. The number  $n_R^i$ represents the number of hypermultiplets in the representation $\textbf{R}$ of the gauge group $G_i$ and $A_R^i,B_R^i,C_R^i$ are the following group theory coefficients:
 	\begin{eqnarray}
 	tr_{\text{R}}F^2=A_R trF^2, \quad	tr_{\text{R}}F^4=B_R trF^4+C_R(trF^2)^2
 	\end{eqnarray} 
the values of those coefficients for various representations and the normalization factors $\lambda_i$ are summarized in \cite{Kumar_2009}.
In addition as shown in \cite{Kumar_2010} the vectors  $a^\alpha,b_i^\alpha \in \R^{1,T}$ are constrained to have integer inner products $ a\cdot a , a\cdot b_i , b_i \cdot b_j \in \Z$ with respect to the bilinear form $\Omega_{\alpha \beta}$, we call this the anomaly lattice. The anomaly lattice as described in \cite{Seiberg_2011} needs to be embedded in the full string lattice of the 6d supergravity.  Moreover, it was shown in \cite{Monnier_2019}  that the vector $a$ is a characteristic vector of the lattice $\Gamma $, meaning that for any $x\in \Gamma$ we have $a\cdot x+x^2\in 2\Z$.

 \textbf{Moduli space consideration:} The moduli space of the 6d $(1,0)$ supergravity locally takes the form $SO(1,T)/SO(T)$ which is parametrized by the a vector $j^\alpha \in \R^{1,T}$ with positive norm $j \cdot j >0$ representing the positivity of the metric on the moduli space.  As discussed in \cite{Kumar_2010,Sagnotti:1992qw} consistency of the theory requires $j \cdot b_i >0, \ j\cdot a<0$. The first set of conditions are required for the positivity of the gauge kinetic terms and the latter condition is associated to the positivity of the Gauss-Bonnet term in gravity \cite{Cheung_2017,Hamada_2019} which has been conjectured to hold. 

 \textbf{BPS string consideration:} The 6D theory has gravity/gauge dyonic strings with charges $a,b_i$. Those charges span the anomaly lattice which is contained in the full string lattice of the 6d theory. Therefore, as discussed in \cite{Seiberg_2011} the anomaly lattice is required to have a unimodular embedding into a self-dual lattice and this fact provides a constraint on possible theories. Furthermore, the existence of the two-form fields $B_2^\alpha$ implies the existence of string sources in accordance with the hypothesis that the spectrum of a gravitational theory needs to be complete \cite{Banks:2010zn,Polchinski:2003bq}. Therefore, according to \cite{Kim:2019aa}, a non-instantonic BPS string with charge Q and non-negative tension  provides the following constraints:
 \begin{eqnarray}\label{uni}
\begin{matrix}
&j\cdot Q\geq 0, Q\cdot Q\geq -1\\ & k_\ell =Q\cdot Q+Q\cdot a+2 \geq 0, \ k_i=Q\cdot b_i\geq 0 \\&  \ \sum_i c_{G_i}\leq c_L=3Q\cdot Q-9 Q\cdot a+2\end{matrix}
 \end{eqnarray}
 where $k_i$ is the level of  $G_i$ and $c_{G_i}$ the central charge associated with the current algebra of $G_i$. In addition, $k_\ell $ is the level of the current algebra associated with $SU(2)_\ell$ which arises from the normal bundle $SO(4)=SU(2)_R\times SU(2)_\ell$ for the  transverse $\R^4$, where $SU(2)_R$  is the R-symmetry of the IR (0,4) SCFT and $SU(2)_\ell$ appears as a left current algebra.

 \textbf{Geometric conditions:} Lastly, we have conditions that arise from string theory considerations and do not seem to have an obvious origin independently. For example,  in F-theory it is required that the vectors $a, b_i$ satisfy the Kodaira conditon\cite{Kumar_2010}:
 \begin{eqnarray}
 j\cdot (-12 a -\sum_i \nu_i b_i)\geq 0 
 \end{eqnarray}
where $\nu_i$ is the multiplicity of the respective singularity or equivalently the number of 7 branes needed  for the non-abelian gauge group $G_i$(e.g. $\nu=N$ for $SU(N)$).
  Additional constraints are imposed from F-theory considerations regarding the irreducibility and effectiveness of divisors. Moreover, for all odd lattices $a$ is primitive and for   $b_i^2\leq 0 $ also $b_i$ is  primitive in F-theory and in that case the former can also be brought to the form $a=(-3,1^T)$ \cite{Kumar_2010}.

\section{Towards a finite Landscape  }
\label{sec:finite Landscape}

\ytableausetup{boxsize=0.7 em,aligntableaux = center}
In \cite{Kumar_2009,Kumar_2010}  it was shown that a large subset of all  possible distinct combinations of non-abelian gauge groups and matter representations that can appear in a 6d ${\cal N }=1$ supergravity is finite for $T<9$.

   \    However, their arguments in some cases  do not generalize to  $T\geq 9$. In particular, \cite{Schwarz_1996,Kumar_2009} provide 5 potentially infinite families with two simple gauge group factors that are not constrained to have an upper bound in the number of massless modes and 3 with three simple gauge factors for $T\geq 9 $. We will argue that those theories are in fact restricted to a finite subset, 
and we will extend the finiteness condition for  more classes of non-abelian theories. 

In this section, we will be making some useful assumptions which we wish to start by justifying. Firstly, as discussed in section \ref{sec:1} a 6d ${\cal N }=1$ supergravity contains two-form fields which could imply the existence of string sources. In particular, the completeness of spectrum hypothesis   will require that all charges compatible with the Dirac quantization condition appear in the theory \cite{Polchinski_2004} and form the string lattice $\Gamma$ of the 6d theory. This statement can be supported from studying black holes \cite{Banks_2011} or in the context of AdS/CFT \cite{harlow2019symmetries}.  Moreover, we can generalize this statement and argue that the lattice of all states should be generated by BPS states because any black hole in the theory could eventually decay to a collection of BPS/anti-BPS states and hence these charges  should be in the lattice too.  Even though this is a heuristic argument, it is a motivation for this assumption.  Therefore we will assume that each charge in $\Gamma$ is a  $\Z$-linear combination of the BPS charges and hence they generate the lattice.  In fact we believe this assumption is more general than the setup we are studying in this paper.  Namely the lattice of allowed BPS charges are generated by BPS generators in all cases.  We are not aware of any counterexamples to this assumption in the string landscape. Therefore, the assumption we will be using can be summarized as follows:
\begin{framed}
	The string charge lattice $\Gamma $ always has a basis of BPS charges that span the entire lattice.
\end{framed}

Secondly, another assumption we will be making is:
\begin{framed}
There are only finitely many inequivalent theories with a given gauge group $G$ and matter $M$.
\end{framed}
Although, we do not have a proof of this statement it constitutes a reasonable physical assumption.  It would be rather strange to have a fixed low energy matter content be represented by infinitely many inequivalent theories.

Lastly, apart from the above assumptions we will also be using the fact that in the case that a particular theory has enough matter to be Higgsed then the string lattice $\Gamma$ of that supergravity does not get affected by the process.  This is because the Higgsing process only involves the hypermultiplets and vector multiplets of the theory and  does not affect the tensor multiplets and hence the dyonic string charge lattice $\Gamma$ should remain unaffected.  In addition, one should note that the  vectors  $a,b_i$ provide the coupling of the two form $B$ fields to the spacetime curvature $B\cdot a trR^2$ and the coupling to the field strengths $B\cdot b_i trF_i^2$. Therefore, since the 6d theory contains strings of charges $b_i$ associated to gauge instantons \cite{Duff_1996} we know that $b_i$ should  belong to $\Gamma$, similarly it has been argued that the vector $a$ should also belong to the lattice corresponding to a gravitational instanton and hence should also be unaffected by the Higgsing process.

We can now move on to constructing  potential infinite families which we wish to exclude. 
Let us recall that in order to construct infinite families of unbounded size one can start by identifying gauge groups that satisfy the $trF^4$ anomalies for arbitrarily large size. 
The simple gauge groups that can have unbounded dimension are $SU(N),SO(N), Sp(N/2)$.
For example, a theory with an $SU(N)$ factor should satisfy:

\begin{eqnarray}
B_{adj}=2N=\sum_R n_R B_R
\end{eqnarray} 

As discussed in \cite{Kumar_2009,Kumar_2010} for large $N$ the only representations that can appear have  $B_R$ at most linear in $N$. Those are  the fundamental, adjoint, two-index antisymmetric and symmetric representations. The set of possible such theories including the groups $SO(N),Sp(N/2)$ is summarized in Table \ref{table:1}.
 \begin{table}[h!]
	\begin{tabular}{|l|l|l|}
		\hline
		Group &Matter & $H-V$\\
		\hline		$SU(N)$ & \begin{tabular}[c]{@{}l@{}} $1\ \text{Adj}$ \\ $1\ \ydiagram{ 2}+ 1 \ \ydiagram{ 1,1}$ \\ $2N\ \ydiagram{ 1}$   \\ $(N+8) \ \ydiagram{ 1}+1 \ $ { \ydiagram{ 1,1}}\\$ (N-8)\ \ydiagram{ 1}+1 \ \ydiagram{ 2}$\\ $16\ \ydiagram{ 1}+ 2 \ \ydiagram{ 1,1}$\\\end{tabular} & \begin{tabular}[c]{@{}l@{}} $ 0 $ \\ $1$ \\ $N^2+1$ \\ ${1\over 2}N^2+{15\over 2}N+1 $ \\ ${1\over 2}N^2-{15\over 2}N+1 $\\ $15N+1$\\\end{tabular}  \\   \hline
		$SO(N)$ & $(N-8)\ \ydiagram{ 1}$             &     $   {1\over 2}                             N^2-{7\over 2}N    $   \\ 
	& \ $1\ \ydiagram{ 1,1}$             &     $  0   $   \\ \hline
		$Sp(N/2)$ & \begin{tabular}[c]{@{}l@{}}$(N+8)\  \ydiagram{ 1} $\\ $16 \ \ydiagram{ 1}+1 \ \ydiagram{ 1,1}$\end{tabular}      &  \begin{tabular}[c]{@{}l@{}}$   {1\over 2}                             N^2+{7\over 2}N  $  \\ $15N-1$\end{tabular}     \\ 
			& \ $1\ \ydiagram{ 2}$             &     $  0   $   \\ \hline
	\end{tabular}
	\caption{Most theories have $H-V\to \infty $ as $N\to \infty $ except those with $H-V=0, 1 $ for which $T\leq 9 $  and there is no obstruction to having an infinitely large gauge group from anomalies alone.}
	\label{table:1}
\end{table}

In particular, we note from Table \ref{table:1} that the only theories that satisfy the gravitational anomaly  for arbitrary $N$ are  $SU(N) \text{ with  }\text{Adj} / 1\ydiagram{2}+1\ydiagram{1,1}$ and $SO(N)/Sp(N/2)$ with $\ydiagram{1,1}/\ydiagram{2} $ with $T\leq 9$.  As discussed in \cite{Kumar_2010}  $T<9$ are excluded since there is no solution for the vectors $a,b$ satisfying $a^2>0,b^2=0, a\cdot b =0$. However, for $T=9$ both $a,b$ are null  vectors with $a\cdot b =0$ and hence parallel, i.e. $b=\lambda a $ with $\lambda <0$ (such that $j\cdot a<0 \text{ and  }j\cdot b>0$). Specifically, in this case it is simple to find solutions  $a,b$ and in fact such an example is constructed later in this section. Therefore, for $T=9$ this theory constitutes  a potentially infinite family with unbounded size.

\begin{table}[h!]
	\begin{tabular}{|l|l|}
		\hline
		$SU(N)\times SU(N)$ &2 (\ydiagram{1},${\ydiagram{1}}$) \\ \hline
		$SO(2N+8)\times Sp(N)$ & $(\ydiagram{ 1},\ydiagram{1})$                                                          \\ \hline
		$SU(N)\times SO(N+8)$ & $(\ydiagram{ 1},\ydiagram{1})$       $+(\ydiagram{ 1,1},1)$                              \\ \hline
		$SU(N)\times SU(N+8)$ & $(\ydiagram{ 1},\ydiagram{1})$       $+(\ydiagram{ 1,1},1)$      $+(1,\ydiagram{ 2})$                         \\ \hline
		$Sp(N)\times SU(2N+8)$ & $(\ydiagram{ 1},\ydiagram{1})$       $+(1,\ydiagram{ 2})$                              \\ \hline
	\end{tabular}
	\caption{Potentially infinite families with two simple gauge group factors.  }
	\label{table:2}
\end{table}

Next we consider theories with gauge groups of the form $G_1\times G_2$ with $G_i$ drawn from Table \ref{table:1}. In \cite{kumar2009string,Kumar_2009,Schwarz_1996}  they identify 5 potentially infinite families with arbitrarily large dimension given in Table \ref{table:2} and composed of two simple gauge factors from Table \ref{table:1}. The expectation is that even though each individual factor may not satisfy the gravitational anomaly, we can arrange such that introducing matter charged under both gauge groups reduces $H-V$ enough to make it possible. Furthermore, it is important to note that the only matter charged under two gauge groups is bifundamental matter. This can be  justified by considering the fact that for 6d ${\cal N}=1$ gauge theories we know that all theories are Higgsable until one reaches the Non-Higgsable Clusters(NHC) \cite{Morrison_2012} or the gauge group gets completely Higgsed away and hence we expect that any family of theories should be Higgsable to some minimal gauge group.  As discussed earlier Higgsing does not affect the string lattice and consequently the vectors $b_i$ of the instantonic strings of the gauge theory, which implies that their inner products defined through  the anomaly cancellations condition (\ref{eqn:F2F2}) should be independent of the size  $N$ of the gauge group which gets reduced by the process.

In particular, for two vectors $b_1,b_2$ their inner product is given by $b_1\cdot b_2=\sum_{R,S}\lambda_i \lambda_j n^{ij}_{RS}A^i_RA^j_S$ which as noted in  \cite{Kumar_2009, Kumar_2010} can only be independent of $N$ if both $R$ and $S$ are the fundamental representations. But more specific to the theories from Table  \ref{table:1}, one can see that no theory has enough matter to gauge any of the $Adj, \ydiagram{2},\ydiagram{1,1}$ because for example a theory of the form $SU(N)\times G_2(N)$ with $(Adj,R )$ would require that $SU(N)$ has $dim(R)$ number of $Adj$ representations but any theory has at most one. Therefore,  no $G_i$ factor can be  $SU(N) \text{ with  }\text{Adj} / 1\ydiagram{2}+1\ydiagram{1,1}$ or $SO(N)/Sp(N/2)$ with $\ydiagram{1,1}/\ydiagram{2} $. 

One could consider $k$ gauge groups from Table \ref{table:1} with matter charged under only one factor and  constant $H-V$. But the gravitational anomaly would then become  $(H_{ch}-V)k\leq 273-29 T$   with $(H_{ch}-V)\geq 0 $ and hence restricting the number of terms. Therefore, we only need to focus on excluding the theories of Table \ref{table:2}\footnote{We note that we do not present the full set of theories since  exchanging matter with its conjugate when gauging it provides distinct theories but this does not affect our calculations.}. The first theory is valid for $T\leq 9 $ and the rest for $T\leq 10$. For $T<9$ it was shown in \cite{Kumar_2010,Kumar_2009} that no solution exists for $a,b_i$ for any of the theories that satisfy all the consistency conditions studied earlier and specifically all $b_i$'s be associated with positive kinetic terms. Similarly, for all theories except the first one,  there  is also no solution for vectors  $a,b_i$ when $T=9$. This is easy to verify for example for  $SO(2N+8)\times Sp(N)$, which has vectors  $a,b_i\in \R^{1,T}$ that satisfy:
\begin{eqnarray}\label{ex1}
a\cdot b_1=2,\  a\cdot b_2=-1, \ b_1^2=-4, \ b_2^2=-1, \ b_1\cdot b_2=2
\end{eqnarray}
There are two   null vectors $a, (b_1+2b_2)$  that satisfy $a\cdot  (b_1+2b_2)=0$ and hence need to be parallel 
$b_1+2b_2=\lambda a \Longrightarrow b_1=\lambda a -2b_2$ for some $\lambda \in \R$. However, since $b_1\cdot b_2=2$ then  $\lambda=0$ implying that  $ j\cdot b_1=-j\cdot 2b_2$, meaning that we can not find vector $j$ ensuring the positivity of both kinetic terms. In an identical fashion one can show the same result by considering the null vectors $a$ and $2b_1+b_2$(for the third)  or $b_1+b_2$ (for the fourth and fifth).

\begin{table}[h!]
	\begin{tabular}{|l|l|}
		\hline
		$SU(N-8)\times SU(N)\times SU(N+8)$ &	$(\ydiagram{ 1}\otimes \ydiagram{ 1}\otimes 1)$ +$(1\otimes  \ydiagram{ 1}\otimes \ydiagram{ 1})$\\ 
		& $+(\ydiagram{ 1,1}\otimes1\otimes 1)+(1\otimes 1\otimes \ydiagram{ 2}) $\\ \hline
	$Sp((N-8)/2)\times SU(N)\times  SO(N+8)$  & $(\ydiagram{ 1}\otimes  \ydiagram{ 1}\otimes 1)$ +$(1\otimes   \ydiagram{ 1}\otimes  \ydiagram{ 1})$              \\ \hline
$SU(N-8)\times SU(N)\times  SO(N+8)$  & $(\ydiagram{ 1}\otimes  \ydiagram{ 1}\otimes 1)$ +$(1\otimes   \ydiagram{ 1}\otimes  \ydiagram{ 1})+(\ydiagram{1,1}\otimes1\otimes1)$       \\ \hline
	$Sp((N-8)/2)\times SU(N)\times  SU(N+8)$  & $(\ydiagram{ 1}\otimes  \ydiagram{ 1}\otimes 1)$ +$(1\otimes   \ydiagram{ 1}\otimes  \ydiagram{ 1})+(1\otimes1\otimes\ydiagram{2})$      \\ \hline
	\end{tabular}
	\caption{Potential infinite families with three simple gauge factors. }
	\label{table:3}
\end{table}
Furthermore, as described in the Appendix  anomalies permit classes of infinite families with more than two simple gauge factors. For example, the  gauge group theories described in Table \ref{table:3} where first introduced in \cite{Kumar_2010}. More generally, one can construct theories which satisfy all the anomalies with arbitrarily large number of gauge factors. Specifically, linear chains of such theories are presented in  Table \ref{table:k}.   In addition, in Appendix \ref{appnon} we discuss theories  that have gauge groups connected in a non-linear fashion.

 For example, one can construct theories where one gauge group is connected to multiple others. Specifically, in \ref{appnon}  we find that a large class of these theories have inner products $b_i\cdot b_j$'s corresponding to the affine ADE, where each $b_i$ represents a node on the Dynkin diagram. Each $b_i$ is associated with a  gauge group  $SU(a^{\vee}_i N )$ where $a^\vee$ is the dual coxeter label and the matter is bifundamentals according to the links of the affine Dynkin diagram. 
 
  However, an interesting observation is that even though anomalies permit the $A,D$ type inner products to have arbitrarily many factors, a more careful analysis shows that this is not possible. This is because the vectors $a, b_i$ form the anomaly lattice  which needs  to  be embedded in the string lattice $\Gamma$. The anomaly lattice is generated by at most $k+1$ vectors $a, b_i$ and the full lattice $\Gamma$ is of signature $(1,(-1)^T)$ and generated by $T+1$ vectors. This implies, that 
 $k\leq T$ and since the gravitational anomaly cancellation requires $T\leq 9 $ then $k\leq 9$. 
 
 Moreover, one can show that $V=\sum_i^k b_i$ is a  null vector which satisfies $a\cdot V=0$. Hence for $T<9$ we have $a^2>0$ which implies that $V=0$ but now one can notice that not  all $j\cdot b_i>0$  can be satisfied simultaneously. Therefore, the only case for which solutions could be found is for $T=9$.

Similarly, for the theories of Table \ref{table:k} with $k$ factors the anomaly lattice has signature $(1,(-1)^{k-1})$ for  $T<8+k$ which means that $k\leq T+1$.  Therefore, using the last column of  Table \ref{table:k} one can show that all the theories have a finite number of gauge factors and more specifically the second theory has $T\leq 138$ while the rest  $T\leq 137$.

Furthermore, we note that the theories of Table \ref{table:k} with $T< 8+k$ have no consistent solutions. In particular, the first and last theories have the same anomaly lattice and hence can be considered together, same holds for the second and third. For the first and last theory one can consider   $T<9$ and   note that $a^2>0$ and $(b_1+\cdots + b_k)^2=0$ with $a\cdot (b_1+ \cdots  b_k)=0$ from which it follows that $b_1+ \cdots  b_k=0$ and hence cannot satisfy $j\cdot b_i>0$ simultaneously for all $i$. We can extend this for $8+k> T\geq 9$ by considering the following vectors:
\begin{eqnarray}
V_1=a+\sum _{i=k-T+9}^{k-1} b_i (i-k+T-8)+(T-9) b_k,	\ V_2=\sum_{i=1}^kb_i
\end{eqnarray}
It is simple to verify that  $V_1^2=V_2^2=V_1\cdot V_2=0 $ from which it follows that $ V_2= \lambda V_1$. Now consider the product 
$\underbrace{b_k\cdot V_2}_{=0}=\underbrace{b_k\cdot  \lambda V_1}_{=\lambda}\Longrightarrow \lambda =0$.
Therefore, $	V_2=0$ and hence not all $j\cdot b_i >0$ conditions can be satisfied. This method though does not constrain the theories that have $T=8+k$ which  arise when $k\leq 6 $ for the first theory and when $k\leq 7 $ for the last and solutions can be found as we will see later. Similarly, one can note that also for the second and third cases there are no solutions for $T< 8+k$.

\begin{table}[h!]
	\hskip-1.0cm
	\scalebox{0.75}{	\begin{tabular}{|l|l|l|}
			\hline
			Gauge group& Matter & Tensors\\\hline
			$SU(N-8)\times SU(N)\times SU(N+8)\times \cdots \times SU(N+8(k-2))$ & $\ydiagram{ 1,1}\otimes1\cdots \otimes 1+1\otimes1 \cdots \otimes  \ydiagram{ 2}$&$T\leq \frac{27 k}{29}+\frac{245}{29}$ \\ \hline
			$Sp((N-8)/2)\times SU(N)\times SU(N+8)\times \cdots \times SO(N+8(k-2))$& &$T\leq \frac{27 k}{29}+\frac{247}{29}$\\ \hline
			$SU(N-8)\times SU(N)\times SU(N+8)\times \cdots \times SO(N+8(k-2))$& $\ydiagram{ 1,1}\otimes1\cdots \otimes 1$ &$T\leq \frac{27 k}{29}+\frac{246}{29} $\\ \hline
			$Sp((N-8)/2)\times SU(N)\times SU(N+8)\times \cdots \times SU(N+8(k-2))$& $1\otimes1 \cdots \otimes  \ydiagram{ 2}$&$T\leq \frac{27 k}{29}+\frac{246}{29}$\\ \hline
	\end{tabular}}
	\caption{Each theory has bifundamental matter between any adjacent groups and the matter indicated in the table is matter charged under only one gauge group. The last column indicates the upper bound on $T$ that the gravitational anomaly imposes.}
	\label{table:k}
\end{table}

We will now provide a general argument that restricts all the theories presented above to a finite set. The argument is based on  \cite{Kim:2019aa} where they use completeness of spectrum as evidence for the  existence of  BPS strings with some charge $Q=(q_1,\cdots q_{10})$ and $q_i\in \Z$ satisfying consistency conditions (\ref{uni}). Those consistency conditions will then provide us with an upper bound on the size $N$ of the gauge group. 

All the theories above have a gauge group with a finite  number of non-abelian simple gauge groups and their size is controlled by the parameter $N$ which is not bounded by the arguments already presented. However, one can notice that each family of theories labelled by $N$ is connected through Higgsing. For example, $SU(N)+1Adj $ can be Higgsed to $SU(N-1)+1Adj $ by making $2N-1  $ full hypermultiplets massive. However, as discussed earlier the Higgsing process does not affect the string lattice which implies that any vectors in the lattice is independent of the size $N$. Specifically, by considering $\{Q_i\}$ as  the BPS string states that generate $\Gamma$  and  satisfy the conditions (\ref{uni}) then one has that these charges are also independent of $N$.
This therefore implies that there should exist a minimal choice of BPS charge $Q\in \{Q_i\}$ that is also independent of $N$. Therefore, for an infinite family drawn from the examples above and using that $Q^2+Q\cdot a \geq -2$  the unitarity bound becomes (for at least one non-zero $k_j$):
\begin{eqnarray}
& k_j{\dim G_j\over k_j+h^\vee_j}\leq c_\ell+\sum_i k_i{\dim G_i\over k_i+h^\vee_i}\leq c_L=3Q^2 -9Q\cdot a+2\leq 12Q^2+20 
\end{eqnarray}
where $k_i$ is the level of the $G_i$ current algebra and $c_\ell ={3k_{\ell}\over 2+k_{\ell}  }$ the central charge of $SU(2)_{\ell}$.
We note that if $k_i\neq 0 $ then for $G_i=SU(N_i) $    we have that $N_i-1={\dim G_i\over 1+h^\vee_i}\leq k_i{\dim G_i\over k_i+h^\vee_i}$, for $G_i=Sp(N_i) $    we have that $2N_i-3\leq {\dim G_i\over 1+h^\vee_i}\leq k_i{\dim G_i\over k_i+h^\vee_i}$, for $G_i=SO(N_i) $    we have that ${N_i\over 2}={\dim G_i\over 1+h^\vee_i}\leq k_i{\dim G_i\over k_i+h^\vee_i}$, where all $N_i$ are a linear function of $N$. This implies that left-hand side of the inequality is always a linear function of $N$'s.
Moreover, since  $Q^2$ is independent of $N$ then this provides a finite upper bound for the size $N$. This is clear if there is one chain of theories related by Higgsing for arbitrarily large $N$.  However, there is a slight loophole in this argument:  it may be that there is no such infinite chain, but that there are infinitely many finite Higgs chains each of which start from a maximal $N_{max}$.  Then by Higgsing them down to a given $N$ we see that for a fixed $N$ we have to have infinitely many inequivalent theories with the same massless matter content which we assumed can never happen.

In the above argument we assumed that at least one of the levels $k_i$ can be chosen to be non-zero.  We now argue that some $Q$ can always be chosen to have at least one non-zero $k_i$.    Let us assume that there is no charge $Q$ such that $b_i\cdot Q=k_i>0$. In this case $b_i\cdot Q=0$ for any of the $Q$'s.   But for any $b_i$ there exists a vector in the lattice which has a non-vanishing inner product with it, by the requirement of the self-duality of the charge lattice.
However, since $Q$ generates the lattice
this leads to a contradiction.  And so there are some BPS states $Q$ with non-vanishing $k_i$.

Even though our argument above does restrict the infinite families to only a finite consistent set under reasonable assumptions, it does not provide us with a concrete upper bound of the size of the gauge groups for each theory. 
We will therefore devote the remaining of the section to go  through some of the theories presented above and find particular solutions for $a,b_i$ such that we can illustrate using unitarity the exact upper bound for the size $N$ in those cases.

Let us begin by considering the single gauge group infinite families:  $SU(N)+1\text{Adj} \text{ or } 1\ydiagram{2}+1\ydiagram{1,1}$ with $T= 9$. In order to ensure that the theory is unitary the following inequality needs to hold:
\begin{eqnarray}
c_\ell+{k(N^2-1)\over k+N}\leq c_L
\end{eqnarray}
where $k$ is the level of the  $SU(N)$ current algebra and $c_\ell ={3k_{\ell}\over 2+k_{\ell}  }$ being the central charge of $SU(2)_{\ell}$. The  inequality would be strongest for a minimal choice of $c_L$ which depends on the choice of charge $Q$. Let us consider a representation of the $a,b$ vectors.  In particular in the integral basis let us take $
	\Omega =\text{diag}(1,(-1)^{9})$, and $ a=(-3, 1^{9}),
$ and we choose a string with minimal charge $Q=(1,-1,-1,0\dots ,0)$ which gives $c_L=8$ and $Q^2=-1, k=-\lambda, k_\ell=0$.  Therefore with this realization, one can easily check that the  only possible $k,N$ that satisfy the unitarity bound are:
\begin{align}
&(k\geq 1, N=0,1,2,3),(4\geq k\geq 1,N=4)\\&(2\geq k\geq 1,N=5),(k=1,N=6,7,8,9)
\end{align}
Therefore, the  size of the gauge group for this theory is bounded by $N\leq 9$ at least for this realization of vectors. In the next section we will show that the theories with $k=1$ belong to the Swampland and hence the size is bounded by $N\leq 5$.  To show that these are general Swampland bounds we need to show that these results hold independently of possible inequivalent realizations of the $(a,b)$ vectors in the lattice. One potential issue is that for $N\leq 3$ there are infinitely many potential solutions for the vector $b$  but  the number of massless modes is still finite. However, in this work, as was discussed earlier, we will assume that there are only finitely many theories with a fixed gauge group and matter, and therefore such issues are avoided.  Furthermore, according to \cite{Kumar_2010}  both vectors $a,b$ need to be primitive in F-theory and hence theories with $\lambda>1$ can not have an F-theory construction. Combining this with our conjecture of the next section we expect that no F-theory construction should be possible also for $\lambda=1$.  

A more general worry is that the above result was deduced with the assumption that $\Omega =\text{diag}(1,(-1)^{9}),a=(-3,1^9)$ while one could imagine other inequivalent choices for these.   In fact, since  $T\equiv1\ (\mod 8)$ in this case we could either have the lattice be odd and isomorphic to $\Z^{T+1}$  or it can be an even lattice isomorphic to $U\otimes E_8(-1)$ with $U=\begin{pmatrix}
0&& 1\\1&&0
\end{pmatrix}$.  One needs to ensure that these as well as other choices for $a$ provide finite size too. Therefore, our previous general argument ensures the finiteness of these theories independently of the type of lattice or particular solution.

Next we move to theories with two simple gauge group factors summarized in Table \ref{table:2}.
\begin{itemize}
	\item $SU(N)\times SU(N)$
\end{itemize}
For  $T=9$ in \cite{Kim:2019aa} it was shown that for a particular choice of $\Omega, a, b_i$'s  all theories with  $N>9$ belong to the Swampland because they contain non-unitary strings. 

More general solutions can be found  by noticing that $a,b_1+b_2 $ are null with  $a\cdot (b_1+b_2 )=0$ and hence satisfy  $-a=m (b_1+b_2)$ with $m>0$.
In this case the general argument translates into the equations $N\leq 12Q^2+20$ with $Q^2$ some constant.

\begin{itemize}
	\item $SO(2N+8)\times Sp(N)$
\end{itemize}
The anomaly cancelation conditions dictate the following inner products between the vectors $a,b_i\in \R^{1,T}$:
\begin{eqnarray}\label{ex}
a\cdot b_1=2,\  a\cdot b_2=-1, \ b_1^2=-4, \ b_2^2=-1, \ b_1\cdot b_2=2
\end{eqnarray}
For $T=10 $  solutions $a,b_i$ exist but we show that unitarity ensures the finiteness of the theory. We may  choose a presentation of these such that the bilinear form $\Omega$ and the vectors $a,b_1,b_2$ are given as follows:
\begin{eqnarray}
\begin{matrix}
\Omega =\text{diag}(1,(-1)^{10}), & a=(-3, 1^{10})\\
b_1=-2 a,  &b_2=(1,-1,-1,0^{8})
\end{matrix}
\end{eqnarray}
In this presentation one can choose $j=(1,0^{10})$ which satisfies $j\cdot a <0$ and $j \cdot b_i>0$ as desired.
Considering a  BPS string with charge $Q=(q_1,\cdots q_{11})$ satisfying conditions (\ref{uni}) then unitarity of the string worldsheet requires that:
\begin{eqnarray}
 {k_1((2N+8)(2N+7)/2)\over k_1+2N+6} + {k_2(2N(2N+1)/2)\over k_2+(N+1)}  \leq c_L 
\end{eqnarray}

One can easily check that a minimal  string charge solution can be $Q=(1,-1,0^8,-1)$, which has levels $k_1=2,k_2=0$ and central charge $c_L=8$. The unitarity bound for this string configuration reduces to:

\[      {2((2N+8)(2N+7)/2)\over 2+2N+6}   \leq 8\Longrightarrow    N\leq 1/2  \]
This seems to be reassuring  because it does not rule out the theories at $	N=0$ with a single $SO(8)$  which do have known string theory realizations\cite{Morrison_2012b, martini20156d,Taylor_2012h}. As for  the case of $N=1/2$  one has a single  $SO(9)$ with 1 fundamental hypermultiplet which is the unHiggsed version of the $SO(8)$ theory and if it exists could have the same base.

\begin{itemize}
	\item 	$SU(N)\times SO(N+8)$ 
\end{itemize}
 We note that the gravitational anomaly restricts $T\leq 10$ and hence we need to ensure finiteness of the theory for $T=10$ as before.
In this case the charge lattice is given by 
\begin{eqnarray}\label{ex4}
a\cdot b_1=-1,\  a\cdot b_2=2,\ b_1^2=-1,  \ b_2^2=-4, \ b_1\cdot b_2=2
\end{eqnarray}

It seems that the anomaly charge lattice is identical to the one before and hence we can use those results. In other words, for  $T=10$  the vectors are identical as in the previous example but with $b_1\leftrightarrow b_2$.
Therefore this string configuration  with $k_2=2, k_1=0$ implies that 

\begin{eqnarray}
{k_2((N+8)(N+7)/2)\over k_2+N+6} \leq 8\Longrightarrow    N\leq 1
\end{eqnarray}

Therefore, as expected this bound  does not rule out the single $SO(8)\text{ or  } SO(9)$ theories as discussed above.
\begin{itemize}
	\item $SU(N)\times SU(N+8)$
\end{itemize}
This family has charge lattice vectors satisfying:
\begin{eqnarray}\label{ex2}
a\cdot b_1=-1,\  a\cdot b_2=1,\ b_1^2=-1,  \ b_2^2=-1, \ b_1\cdot b_2=1
\end{eqnarray} 
Similarly to the previous example for $T=10$ such vectors exist but there are finitely many consistent unitary solutions.  One such representation is given by the choice
\begin{eqnarray}\label{vecs1}
\begin{matrix}
\Omega =\text{diag}(1,(-1)^{10}), & a=(-3, 1^{10})\\
b_1=(1,-1,-1,0^8),  &b_2=-a
\end{matrix}
\end{eqnarray}
One can easily check that $Q=(1,-1,0^8,-1)$  is a minimal string charge which satisfies eq.(\ref{uni}) 
with levels $k_1=0,k_2=1$ and $c_L=8$. For the string configuration to be unitary we need to satisfy:
\begin{eqnarray}
  {((N+8)^2-1)\over 1+(N+8)}  \leq 8\Longrightarrow    N\leq 1  
\end{eqnarray}
This bound potentially allows for $N=0,1$ corresponding to $SU(8)+\ydiagram{2}$ and $SU(9))+\ydiagram{1}+\ydiagram{2}$. However, such string theory realizations are not known and as we will argue in the next section these theories belong to the Swampland.

\begin{itemize}
	\item $Sp(N)\times SU(2N+8)$
\end{itemize}This theory has the same anomaly lattice as  (\ref{ex4})  and hence we can reuse those results. For $T=10$ vectors $a,b_i$ can be found as in (\ref{vecs1}). Therefore, for $k_1=0,\ k_2=1$ we see that unitarity implies 
\begin{eqnarray}
{k_2((2N+8)^2-1)\over k_2+2N+8} \leq 8\Longrightarrow    N\leq 1/2 
\end{eqnarray}
This bound allows for $N=0,1/2$ corresponding to $SU(8)+\ydiagram{2}$ and $SU(9)+\ydiagram{1}+\ydiagram{2}$ but as was discussed these theories will be ruled out in the next section. 

More generally, since for the last two examples $(a+b_2),(b_1+b_2)$ are null and orthogonal, the most general vectors needed are given by the family of solutions $a=\lambda \ b_1+(\lambda-1)b_2$ with $\lambda\leq 0$ in order to ensure positivity of the kinetic terms(for the first two examples one can replace $ b_1\to 2b_1$). 
Similarly to the first example, since  the above theories can be  Higgsed from $N$ to $N-1$ then unitarity would imply the finiteness of each family of theories as was discussed earlier.

Next we  move on to theories with three simple gauge factors. For example, the set of theories from Appendix \ref{appnon} have  $b_i$'s form inner products according to the affine ADE algebras. For example, the $\hat{A}_2$ type theory with $SU(N)^3$ and  $T=9$ has the anomaly lattice:

\begin{eqnarray}\label{anlat}
\Lambda=\left(
\begin{array}{cccc}
a^2& -a\cdot b_1& -a\cdot b_2&-a\cdot b_3 \\
-a\cdot b_1&b_1^2 & b_1\cdot b_2 & b_1\cdot b_3 \\
-a\cdot b_2&b_1\cdot b_2 & b_2^2 & b_2\cdot b_3\\
-a\cdot b_3&b_1\cdot b_2 & b_2\cdot b_2 & b_3^2 \\
\end{array}
\right)=
\left(
\begin{array}{cccc}
0& 0& 0&0 \\
0&-2 & 1 & 1 \\
0&1 & -2 & 1 \\
0&1 & 1& -2 \\
\end{array}
\right)
\end{eqnarray}
These inner products can be solved for vectors satisfying the linear relation $a=\lambda(b_1+b_2+b_3)$ for $\lambda<0$.
 For example, a solution to the anomaly  lattice (\ref{anlat}) is given by:
\begin{eqnarray}
\begin{matrix}
\Omega =\text{diag}(1,(-1)^{9}), & a=(-3, 1^{9})\\
b_1=(1,-1,-1,-1,0^6),  &\quad b_2=(1,0^3,-1,-1,-1,0^3),& \quad b_3=(1,0^6,-1,-1,-,1)
\end{matrix}
\end{eqnarray}
One can choose $j=(1,0^9)$ and charge $Q=(1,-1,0,0,-1,0^5)$ which gives $k_1=k_2=0$ , $k_3=1$ and $c_L=8$. Therefore, worldsheet unitarity implies:
\begin{eqnarray}\label{sun3}
{(N^2-1)\over 1+N}\leq 8 \Longrightarrow N\leq 9
\end{eqnarray}

Therefore, for the particular  choice of anomaly vectors this theory is finite and theories with $N>9$  belong to the Swampland. For more general possible representations  of the vectors, the argument works exactly as discussed earlier.  

Furthermore, other types of theories with three gauge groups can be found in  Table \ref{table:3} but they are all particular cases of those in Table \ref{table:k} for $k=3$ and hence can be handled together. 
\begin{itemize}
	\item $SU(N-8)\times SU(N)\times SU(N+8)\times \cdots \times SU(N+8(k-2))$
\end{itemize}

For  this theory the maximum number of tensor multiplets  arises for $k\leq6$ and has   $T_{max} = 8+k$ and these constitute the only case we need to consider as the other values of $T$ were ruled out earlier. The anomaly  charge lattice of the strings determined by the type of gauge group and matter is given by :

\begin{align}\Lambda=\left(
\begin{matrix}
a^2 & -a\cdot b_1 & -a\cdot b_2& -a\cdot b_3 &\cdots & -a\cdot b_k \\
-a\cdot b_1 & b_1^2 & b_1\cdot b_2 & b_1\cdot b_3 &\dots & b_1\cdot b_k \\
-a\cdot b_2& b_1\cdot b_2 & b_2^2& b_2\cdot b_3  &\dots & b_2\cdot b_k \\
\vdots &\vdots & \vdots &\vdots & \ddots &\vdots \\
-a\cdot b_k& b_1\cdot b_k & b_2\cdot b_k  & b_3\cdot b_k &\dots & b_k\cdot b_k 
\end{matrix}
\right)=
\left(
\begin{matrix}
9-T & 1 & 0 & 0 & \cdots  & -1 \\
1 & -1 & 1 & 0 & \cdots  & 0 \\
0 & 1 & -2 & 1 & \dots & 0 \\
0 & 0 & 1 & -2 & \cdots  & 0 \\
\vdots & \vdots & \vdots &\vdots &\ddots & \vdots\\
-1 & 0 & 0 & 0 & 1 & -1 \\
\end{matrix}
\right)
\end{align}

We may  consider a particular solution for the vectors $a,b_i$ given by:
\begin{eqnarray}
\begin{matrix}
a=(-3,1,1,1\cdots, 1),&b_1=(1,-1,-1,0\cdots, 0),&b_2=(0,0,1,-1\cdots, 0)\\
b_3=(0,0,0, 1,-1\cdots, 0),&b_{i}=(0\cdots, 1,-1,\cdots, 0),& b_k =-a+\sum _{i=1}^{k-2} b_i (-i+k-1)\\
\end{matrix}
\end{eqnarray}
Moreover, we also need to identify a consistent K\"ahler form $j$ and we would like to make a minimal choice of string charge $Q$ for each $k$.

For $k=3 $ we may choose: $j=(2,0,0,1,0^8), Q=(1,-1,0^9,-1)$
for which $k_1=0,k_2=0,k_3=1,c_L=8$. Therefore, string unitarity can be expressed as:
\begin{eqnarray}
   {k_3((N+8)^2-1)\over k_3+N+8} \leq 8\Longrightarrow    N<=1 
\end{eqnarray}

For $k=4 $ we may choose: $j=(  3,0,0,1,2,0^8  ),Q=(1,-1,0^{10},-1)$
with $k_1=0,k_2=0,k_3=0,k_4=1,c_L=8$
\begin{eqnarray}
  {k_4((N+8(4-2))^2-1)\over k_4+N+8(4-2)} \leq 8\Longrightarrow    N\leq -7
\end{eqnarray}

For $k=5 $ we may choose: $j=(4,0,0,1,2,3,0^8),Q=(1,-1,0^{10},-1)$
with $k_1=0,k_2=0,k_3=0,k_4=0,k_5=1,c_L=8$
\begin{eqnarray}
 {k_5((N+8(5-2))^2-1)\over k_5+N+8(5-2)} \leq 8\Longrightarrow    N\leq -15 
\end{eqnarray}

For $k=6 $ we may choose: $j=(6,0,0,1,2,3,4,0^8),Q=(1, -1, 0^{12}, -1)$
with $k_1=0,k_2=0,k_3=0,k_4=0,k_5=0,k_6=1,c_L=8$.
\begin{eqnarray}
   {k_5((N+8(6-2))^2-1)\over k_5+N+8(6-2)} \leq 8\Longrightarrow    N\leq -23  
\end{eqnarray}

We therefore, conclude that the above inequalities suggest that only $SU(9)+1\ydiagram{1}+1\ydiagram{2}$ and  $SU(8)+1\ydiagram{2}$ are allowed which as was discussed earlier will be ruled out in the next section.

\begin{itemize}
	\item $
Sp((N-8)/2)\times SU(N)\times SU(N+8)\times \cdots \times SO(N+8(k-2))
$
\end{itemize}
The anomaly lattice is the same as in the previous theory except with $b_k\to2 b_k$ and  the maximum $T_{max}=k+8$  attained for $k\leq 7$.

For $k=3$ we have $k_1=0,k_2=0,k_3=2,c_L=8$ giving us:
\begin{eqnarray}
    {2((N+8)(N-7)/2)\over 2+N+6} \leq 8\Longrightarrow    N\leq 1  
\end{eqnarray}

For $k=4$ we have $k_1=0,k_2=0,k_3=0,k_4=2,c_L=8$ giving us:
\begin{eqnarray}
  {2((N+8(4-2))((N+8(4-2)-1)/2)\over 2+N+8(4-2)-2} \leq 8\Longrightarrow    N\leq -7 
\end{eqnarray}

For $k=5$ we have $k_1=0,k_2=0,k_3=0,k_4=2,k_5=2,c_L=8$ giving us:
\begin{eqnarray}
    {k_5((N+8(5-2))^2-1)\over k_5+N+8(5-2)} \leq 8\Longrightarrow    N\leq -15 
\end{eqnarray}

For $k=6$ we have $k_1=0,k_2=0,k_3=0,k_4=0,k_5=0,k_6=2,c_L=8$ giving us:
\begin{eqnarray}
 {k_6((N+8(6-2))^2-1)\over k_6+N+8(6-2)} \leq 8\Longrightarrow    N\leq -23
\end{eqnarray}
For $k=7$ we $j=(8,0,0,1,2,3,4,5,0^8),Q=(1, -1, 0^{13} -1)$ with 
\begin{eqnarray}
k_1=0,k_2=0,k_3=0,k_4=0,k_5=0,k_6=0,k_7=2,c_L=8
\end{eqnarray}
\begin{eqnarray}
   {k_7((N+8(7-2))^2-1)\over k_7+N+8(7-2)} \leq 8\Longrightarrow    N\leq -31 
\end{eqnarray}
Therefore, unitarity implies that the only theories that survive are $SO(9)+1\ydiagram{1}$ or $SO(8)$ which have been discussed earlier in this section. 

Finally, there are two more infinite families of this type that can be found by replacing  $Sp\to SU$ or $SO\to SU$ giving us identical results to those above. 

\begin{itemize}
	\item  $SU(N)^k$ 
\end{itemize}
 Earlier in this section the cases $k=2,3$ where shown to be finite and hence we need to focus on $k>3$ for $T=9$. 
The anomaly lattice of this theory is determined by the inner products:
\begin{eqnarray}
a^2=0, a\cdot b_i=0, b_i \cdot b_{i+1}=1,b_i^2=-2, b_1\cdot b_k=1
\end{eqnarray}

Consider the quadratic form to be $\Omega=\text{diag}(1,(-1)^9)$ then a solution to the anomaly lattice for $k\leq 9$ (the upper bound was determined by requiring the anomaly lattice to embed into $\Gamma$) is:
\begin{eqnarray}
\begin{matrix}
a=(-3,1,1,1\cdots, 1),&b_1=(0,1,-1,0\cdots, 0),&b_2=(0,0,1,-1\cdots, 0)\\
b_3=(0,0,0, 1,-1\cdots, 0),&b_{i}=(0\cdots, 1,-1,\cdots, 0),& b_k =-a-\sum _{i=1}^{k-1} b_i \\
\end{matrix}
\end{eqnarray}
A compatible K\"ahler  form can be found for example $j=(4,1,2,\cdots ,9)$.
For $k<9$  a minimal choice of BPS string charge  is $Q=(1,-1,0\cdots, 0 ,-1)$ which has $Q^2=-1$ and $Q\cdot a=-1$ and $k_1=1,k_{i}=0, k_k =0$ and hence 
\begin{eqnarray}
{(N^2-1)\over 1+N }\leq c_L=8\Longrightarrow N\leq  9
\end{eqnarray}
which is the same result we found previously for $k=3$.

For  $k=9$ a minimal choice of BPS string charge  is $Q=(1,-1,0\cdots, 0 ,0)$ which has $Q^2=0$ and $Q\cdot a=-2$ and $k_1=1,k_{i}=0, k_k =1$ and hence 
\begin{eqnarray}
{2(N^2-1)\over 1+N }\leq c_L=20\Longrightarrow N\leq  11
\end{eqnarray}

As presented earlier in the section and in the Appendix one can see that there are more theories that we could analyze but the methods are parallel to those already discussed. Therefore, the general argument in the beginning of the section applies to those infinite families too and similar choice of solutions as to those already made would reveal potential upper bounds for the sizes of the gauge groups.

We note that our general argument resticted the dimension of each gauge group to be finite. Additionally, we were able to show that a number of theories with $SU(N),SO(N),Sp(N/2)$ type gauge groups may only have finitely many  simple gauge groups by studying the lattice embedding of the anomaly lattice to the full 6d string lattice. However, more theories can be constructed with bounded dimension and unbounded number of tensor multiplets allowed by anomalies.

Recall that the gravitational anomaly is given by 
\begin{eqnarray}\label{gr2}
H_{ch}-V\leq 273-29 T
\end{eqnarray}
As we have seen before constructing  theories with arbitrarily many gauge factors not restricted by anomalies requires $H-V<0$ so that eq. (\ref{gr2}) is always satisfied. Therefore, if one could choose more theories of finite dimension and minimal matter that satisfy the anomaly conditions but have negative $H-V$ then it could be possible to have an unbounded number of those. Additionally,  assuming that $H_{ch}-V<0$ for a given simple gauge group we can rearrange eq. (\ref{gr2}) to write it as 
\begin{eqnarray}
T\leq {273\over 29}-{(H_{ch}-V)k\over 29}
\end{eqnarray}
where $k$ is the number of simple gauge factors.
However, as was discussed earlier in this section one needs to be able to embed the anomaly lattice in the full string lattice of the 6d theory and hence satisfy $k\leq T$. This is possible only if $(H_{ch}-V)\leq -29$.

Examples of theories with minimal matter include the NHC's and more found in \cite{Morrison_2012, Heckman_2019}. For example, pure $SO(8)$ has $H_{ch}-V=-28$  and $SO(9)+1\ydiagram{1}$ has $H_{ch}-V=-27$. However, neither satisfy $H_{ch}-V\leq -29$ and hence one can not have an infinite number of those.
Likewise, also $SU(3)^k$ is bounded because $H_{ch}-V=-8$ and hence $k\leq 17$. Also, for $(g_2\times SU(2))^k$ one has $H_{ch}-V=-9$ and for $(SU(2)\times SO(7)\times SU(2))^k$ one has $H_{ch}-V=-11$. 
Therefore, from the NHC's the following are compatible with $(H_{ch}-V)\leq -29$:

\begin{itemize}
	\item $f_4$ with $b_i\cdot b_i=-5$.
	
	The gravitational anomaly determines that $T\leq\frac{52 k}{29}+\frac{273}{29} $. For example when $T=k+9$ we can find solutions of the form:
	\begin{align}
	\hskip -1 cm 
	&	a=(-3,1^T)\\
	& b_{1}=(-1,-1,-1,2,0^{T - 3 }) \\ & b_{2}=(0,0,-2,-1,0^{T - 3  })  \\
	& \vdots \\
		&	b_{i}=(-1,-1,0^{2(i-1)},-1,2,0^{T - 1- 2 i })	\\ & b_{i+1}=(0^{2{	i}},-2,-1,0^{T - 1- 2 {	i }}) \\
		& \vdots \\
		&	b_{k-1}=(-1,-1,0^{2(k/2-1)},-1,2,0^{T - 1- 2 	k/2})	\\ & b_{k}=(0^{2	k/2},-2,-1,0^{T - 1- 2 {	k/2 }})
	\end{align}
	
If $k$ is odd just replace $k/2\to \lfloor k/2\rfloor$ and $k\to k-1$ and add as the last vector: $	b^{odd}_{k}=(-1,-1,0^{2(\lfloor{k/2}\rfloor)},-1,2,0^{T - 3- 2 \lfloor{k/2}\rfloor })	$. As for K\"ahler class we can choose:
\begin{eqnarray}
j=(-j_0,1^T) \text{ for } {T\over 3}\geq j_0>\sqrt{T }  \ e.g \ j_0 =\lfloor{k/3}\rfloor -1 \text{ and } k\geq 21
\end{eqnarray}
where the upper bound is chosen such that $-j\cdot a >0$ and the lower bound to ensure $j^2>0$. Moreover, it is also simple to check that $j\cdot b_i>0.$
One could find more solutions for small $k$ but we are only interested in this work for large $k$ and hence we will not attempt to enumerate those.
This choice of vectors shows that anomalies permit to have unbounded many such gauge groups. 

However, one could consider  a string  with minimal charge $Q=(-q,0^T)$.	This choice of charge has: $k_{i=odd}=q,k_{i=even}=0$, $k_\ell=q^2+3q+2\geq 0, c_R=3q^2-9q\geq 0$ true for $q \geq 3 $. However imposing worldsheet unitarity 
	\begin{eqnarray}
	{3(q^2+3q+2)\over 2+(q^2+3q+2) }+	\lceil{k\over 2}\rceil{52\over q+9}\leq  3q(q-9)+2
	\end{eqnarray}
	one can note that the inequality cannot be satisfied when $3\leq q\leq 9$ for any $k$(with lower bound as discussed above).
	
	In particular, more generally  these solutions are valid for any $k+2\leq T\leq\frac{52 k}{29}+\frac{273}{29} $ and hence similarly restrict $k$ just as we saw above. 
	\item $e_6$ with $b_i\cdot b_i=-6$
	
	The gravitational anomaly imposes that $T\leq \frac{78 k}{29}+\frac{273}{29}$. For example a solution can be found when $T=2k+9$:
	\begin{align}
	\hskip -1 cm 
	&	a=(-3,1^T)\\
	&	b_{1}=(-1,-1,1,-2,1,0^{T - 4  }) \\ & b_{2}=(0^3,-1,-2,-1,0^{T-5 })\\
	& \vdots \\
	&	b_{i}=(-1,-1,0^{4(i-1)},1,-2,1,0^{T - 4 i }) \\ & b_{i+1}=(0^3,0^{4(i-1)},-1,-2,-1,0^{T-4 i-1 })\\
	& \vdots \\
	&	b_{k-1}=(-1,-1,0^{4(k/2-1)},1,-2,1,0^{T - 4 k/2 }) \\ & b_{k}=(0^3,0^{4(k/2-1)},-1,-2,-1,0^{T-4 k/2-1 })
	\end{align}
If $k$ is odd as before we can replace $k/2\to \lfloor k/2\rfloor$ and $k\to k-1$ and add as the last vector: $	b^{odd}_{k}=(-1,-1,0^{4(\lfloor k/2\rfloor)},-1,-2,-1,0^{T - 4 (\lfloor k/2\rfloor+1) })	$. As for K\"ahler class we can choose: $ j=(-j_0,1^T) \text{ for } {T\over 3}\geq j_0\geq \sqrt{T }  $.

	 However, just as we saw above strings with charge $Q=(-q,0^T)$ and $3\leq q<10$ satisfy $c_R\geq 0, k_\ell\geq 0 $ but are none unitary because unitarity relation cannot be satisfied:
	\begin{eqnarray}
	{3(q^2+3q+2)\over 2+(q^2+3q+2) }+		\lceil{k\over 2}\rceil{78\over q+12}\leq  3q(q-9)+2
	\end{eqnarray}
	
	More, generally these solutions can be adjusted and used for any 		$ 2k+2\leq T$. 
	Apart from the NHC one can note  that also $e_6 $ with $1$ fundamental hypermultiplet is possible \cite{Heckman_2019}. This theory has $b_i^2=-5, a\cdot b_i=3$ and $-51 k \leq 273-29T$. The analysis of this is very similar to $f_4$ above so we will not repeat it.
	\item $e_7$ with $b_i\cdot b_i=-7$ with ${1\over 2}56$ matter
	
	The gravitational anomaly imposes that $T\leq \frac{105 k}{29}+\frac{273}{29}$. For example a solution can be found when $T=3k+9$:
	\begin{align}
	\hskip -1 cm 
	&	a=(-3,1^T)\\
	&	b_{1}=((-1)^2,-1,-2,(1)^2,0^{T-5}),	 \\& b_{2}=(0^2,(-1)^2,-2,-1,0^{T-5 }) \\& \vdots \\
	&	b_{i}=((-1)^2,0^{5(i-1)},-1,-2,(1)^2,0^{T-5 i }),	 \\& b_{i+1}=(0^2,0^{5(i-1)},(-1)^2,-2,-1,0^{T-5 i })
	\\& \vdots \\
	&	b_{k-1}=((-1)^2,0^{5(k/2-1)},-1,-2,(1)^2,0^{T-5 k/2 }),	 \\& b_{k}=(0^2,0^{5(k/2-1)},(-1)^2,-2,-1,0^{T-5 k/2 })
	\end{align}
	Similarly, as  above strings with charge $Q=(-q,0^T)$ where $3\leq q<10$ are none unitary because they do not satisfy unitarity relation:
	
	\begin{eqnarray}
	{3(q^2+3q+2)\over 2+(q^2+3q+2) }+		\lceil{k\over 2}\rceil{133\over q+18}\leq  3q(q-9)+2
	\end{eqnarray}

	These solutions can be used for any $T$ such that 		$3k+1\leq T$ giving the same result.
	\item $e_7$ with $b_i\cdot b_i=-8$ 
	
	The gravitational anomaly imposes that $T\leq \frac{133 k}{29}+\frac{273}{29}$. For example the following solutions can be found when $T=4k+9$:
	\begin{align}
	\hskip -1 cm 
	&	a=(-3,1^T)\\
	&	b_{1}=((-1)^2,(-1)^2,-2,(1)^2,0^{T - 6  })	\\ & b_{2}=(0^3,(-1)^3,-2,-1,0^{T-7 })\\ &\vdots \\
	&	b_{i}=((-1)^2,0^{6(i-1)},(-1)^2,-2,(1)^2,0^{T - 6 i })	\\ & b_{i+1}=(0^3,0^{6(i-1)},(-1)^3,-2,-1,0^{T-6 i-1 })
	 \\ & \vdots\\
	&	b_{k-1}=((-1)^2,0^{6(k/2-1)},(-1)^2,-2,(1)^2,0^{T - 6 k/2 })	\\ & b_{k}=(0^3,(-1)^{6(k/2-1)},-2,-1,0^{T-6 k/2-1 })
	\end{align}
	For the stings of charge  $Q=(-q,0^T)$ one has $k_{i=odd}=q,k_{i=even}=0$, $k_\ell \geq 0, c_R\geq 0$ true for $q \geq 3 $ but the unitarity bound:
	\begin{eqnarray}
	{3(q^2+3q+2)\over 2+(q^2+3q+2) }+		\lceil{k\over 2}\rceil{133\over q+18}\leq  3q(q-9)+2
	\end{eqnarray}
	shows that strings with $3\leq  q\leq 10$ are non-unitary. Generically, we can find such solutions for all $3k+3\leq T$.
	
		Apart from the two NHCs we studied  one can note that also $e_7 $ with $1,{3\over 2}$ fundamental hypermultiplet are possible. These theories have $b_i^2=-6/-5, a\cdot b_i=4/3$ and $-77 k /-49 k \leq 273-29T$ respectively. The analysis of this is very similar to $f_4,e_6$ as above and hence we will not repeat. 
		
	\item $e_8$ with $b_i\cdot b_i=-12$ 
	
	A specific solution for this theory for large $T$ is discussed in \cite{Kumar_2010, Kim:2019aa} where in the latter work they show that $k$ can not be arbitrarily large for that solution.
\end{itemize}
Even though in the last four cases we do not have a more general way to show that there can only be finitely many terms, the solutions above seem to suggest so.

To sum up, in this section we have shown that certain theories which could potentially be allowed to have arbitrarily large size or arbitrarily many gauge factors, have in fact an upper bound or a more careful analysis reveals that they do not exist. 

This gives a positive answer to the assumption of the Lampost principle that there should be an upper bound on the number of massless modes in a theory of quantum gravity at least for the majority of the proposed infinite families of anomaly free matter content.

\section{A bound on the matter representations}
 \ytableausetup{boxsize=0.7 em,aligntableaux = center}
 In this section we will propose further consistency conditions that need to be imposed for a consistent 6d supergravity theory. In the previous section we summarized how  the existence of BPS strings strongly constrains the bulk theory. In particular, the bulk gauge groups emerge as current algebras on the 2d worldsheet and together with unitarity on the worldsheet, one can impose constraints on the rank of the gauge groups. Such techniques were used in \cite{Kim_2020d} to put an upper bound on the rank of gauge groups for all supergravity theories with 16 supercharges. Moreover, BPS strings imposed constraints on theories with 8 supercharge   in 5d \cite{Katz_2020} and 6d \cite{Lee_2019, Kim:2019aa}, for abelian and non-abelian theories respectively. Moreover, we extensively used such techniques in the previous section. Here we will introduce another constraint that the 6D theory needs to satisfy associated with consistency of the 2d worldsheet with specific type of bulk matter.

  In particular, we now argue that massless matter hypermultiplets in the bulk correspond to relevant/marginal vertex operators on the string. Evidence to support this claim comes from the fact that at least, when giving a vev to a charged massless hypermultiplet it can Higgs the bulk gauge group,  the worldsheet theory of the BPS string for which there is a flavor current associated to the group should get deformed.  This is because the gauge symmetry in the bulk induces the flavor symmetry on the BPS string and consequently the Higgsing process also reduces the flavor symmetry on the BPS string.  This means that there must exist a relevant/marginal deformation of the BPS worldsheet associated to a primary field in representation $\textbf{R}$ of the matter field on the worldsheet (note that non-primary fields except from the current itself will always have dimension bigger than 1).
 Since the current is on the left-moving sector of the string which is non-supersymmetric, this means that there is an operator of left-moving dimension less than or equal to 1 associated to a primary field of representation $\textbf{R}$.  This argument can be extended to all massless representations regardless of whether they can Higgs the gauge group:  Having  massless fields in the representation $\textbf{R}$ of a gauge group should lead to at most marginally irrelevant deformations.  In other words giving a vev to them is obstructed by more than quadratic terms in the bulk theory.  So at the quadratic/leading level they behave as if they are Higgsing the bulk theory and so should be at most marginally irrelevant, i.e. dimension no more than 1.
 
 A simple example of this condition is realized in the heterotic string on $K3$, where the massless charged fields are represented by primary fields with (left,right) dimension $(1,1/2)$ of the $(0,4)$ supersymmetric theory on the worldsheet.
   
To summarize  we have argued that the hypermultiplets transforming in  a particular  representation $\textbf{R}$ need to satisfy the following conditions: 
 \begin{framed}
 	
 \begin{enumerate}
 	 \item The vertex operator of the massless modes with representation $\textbf{R}$ of $G$ with conformal weight $\Delta_ \textbf{R}= {C_2(\textbf{R})\over 2 (k+h^v)}$ where $C_2(\textbf{R})$ is the second Casimir of the $\textbf{R}$ must obey:
 	\begin{eqnarray}
 	\Delta_ \textbf{R}\leq 1
 	\end{eqnarray}
 	\item The representation  $\textbf{R}$ of a primary  with highest weight $\mathbf{\Lambda}=(\Lambda_i,\cdots,\Lambda_{r})$ where $r$ is the rank of the Lie algebra must satisfy :
 	\begin{eqnarray}\label{condition1}
 	\sum_i^r \Lambda_i \leq k 
 \end{eqnarray}
 where $k$ is the level of the current algebra of G on the worlsheet.

 \end{enumerate}
\end{framed}
The first  condition as discussed above requires the  hypermultiplet states of the spacetime theory to  appear as vertex operators in the WZW model and in particular they need to be relevant/marginal primary fields. Therefore, the conformal dimension associated to the hypermultiplets can be at most $1$.
 The second condition is a standard result of the highest-weight representation in Kac-Moody algebras \cite{DiFrancesco:1997nk}. 
 
In addition, these inequalities are independent of the dimension of spacetime and can also be extended to BPS strings in 5d and 4d. For example, in 5d $N=1$ we have monopole strings which need to satisfy the above consistency conditions in the presence of bulk matter and hence constraining the possible representations that can appear.

For example, consider the 5d $N=1$, $SU(2)\times U(1) $  theory constructed in \cite{Katz_2020} with the geometry being the singular quintic with $A_1$ singularity along a curve of degree $d$ and genus $g$. Assuming that $H$ is the proper transform of the hyperplane class of the
quintic, and E the exceptional divisor of the blowup then the following relations are true:
\begin{eqnarray}
H^3=5,H^2 E=0,HE^2=-2d,E^3=4-4g-5d
\end{eqnarray}

In this case the t'Hooft anomaly of the non-abelian gauge symmetry is given by:
\begin{eqnarray}
{-1\over 4 }k_i trF_i^2
\end{eqnarray}
with $k_i=-h_{i,a}q^a$, where $h_{i,a}$ are the the coefficients in the gauge coupling $h_i $ for $G_i$ in the bulk effective action and $q^a$ the string charges. Therefore,  the levels  of  U(1)  , SU(2) with  divisors $H,E$ respectively and $q=(1,0)$ are:
\begin{eqnarray}
k_0 =C_{000}=H^3, k_1=-{3\over 6 } C_{011}={-3\over 6}HE^2=d
\end{eqnarray}
which implies that condition (\ref{condition1}) is given by:
\begin{eqnarray}
\sum_i \Lambda_i\leq d
\end{eqnarray}

Therefore, for a degree $d=1$ curve we can only have fundamental matter in $\textbf{2}$ of $SU(2)$.
This is in accordance with the fact that 
\begin{eqnarray}
E^3=4-4g-5d=-1 \text{\ for\   }d=1,g=0
\end{eqnarray}
was interpreted as having $N=9$ fundamental hypers rather than $1 $ adjoint and $1$ fundamental since the genus was zero. Geometrically, this is the fact that  there is no degree 1 genus 1 curve.

However, if $d=2$ we have 
\begin{eqnarray}
\sum_i \Lambda_i\leq 2
\end{eqnarray}
and say $E^3=-6$ could be either $N=14$ fundamental hypers or $N=6$  fundamental hypers and 1 adjoint. In other words our inequality does not restrict which case it is. From geometry we know that the first case is  correct  in this  example because $E^3=4-4g-5d=-10$ for a genus 1 and degree 2 curve.

Returning to 6d we are interested in seeing how these inequalities can help us as Swampland conditions. 
Let us start by considering the 6d supergravity theory coupled to 	$SU(N)$ with $ (N-8)\ \ydiagram{ 1}+1 \ \ydiagram{ 2}$. The gravitational anomaly restricts these theories to exist up to $T=10$ and the gauge/gravitational anomalies are cancelled for $a\cdot b =1, b \cdot b =-1$.  
We can  choose a basis such that the bilinear form and the vectors $a, b$ are given by:
\begin{eqnarray}
\Omega =\text{diag}(1,(-1)^T),\ a=(-3,1^{T}), \ b=(0^{T},-1) \ 
\end{eqnarray}
In this particular basis we can choose the K\"ahler form to be $J=(n ,0^{T-1},1)$ which satisfies $J^2\geq 1$ for $n\geq 1$ and $J\cdot a < 0 , J \cdot b > 0 $. Now we consider a BPS string with charge $Q=(q_0,\cdots q_T )$ which  must satisfy eq.(\ref{uni}):
\begin{eqnarray}
q_0^2-\sum_{i=1}^Tq_i^2\geq -1, \ q_0^2-\sum_{i=1}^Tq_i^2-3 q_0 -\sum_{i=1}^Tq_i\geq -2, k =Q\cdot b \geq  0
\end{eqnarray}
A string charge consistent with these inequalities is  $Q=(3,0^{T-1},1)$ which gives level $k=1$ for any $T$.
We can now use  eq.(\ref{condition1}) which states that every representation should satisfy:
\begin{eqnarray}
\sum_i \Lambda_i \leq k =1
\end{eqnarray}
However, the symmetric representation has highest weight $\Lambda=(2,0^{N-2})$ and therefore does not satisfy this inequality. We conclude that this theory belongs to the Swampland. This is consistent with the observation in \cite{Kumar_2010}  that for $T=1$ this theory has no F-theory realization.

Another example, that was also discussed in the previous section is:
$SU(N)+1\textbf{Adj} \text{ or } 1\ydiagram{2}+1\ydiagram{1,1}$ with $T= 9$. We found that the  following choices of $(k,N)$ are consistent by using unitarity considerations:
\begin{align}
&(k\geq 1, N=0,1,2,3),(4\geq k\geq 1,N=4)\\&(2\geq k\geq 1,N=5),(k=1,N=6,7,8,9)
\end{align}
However, if we apply condition  (\ref{condition1}) we see that $k=1$ is not a consistent choice because both the adjoint and symmetric representation have $\sum_i\Lambda_i=2$. Therefore in particular all theories with $N>5$ belong to the Swampland.
 \ytableausetup{boxsize=0.3 em,aligntableaux = center}

Consequently, the second condition has helped us rule out theories that do not have string theory realizations but methods such as unitarity bounds of the previous section did not exclude them. However, the first condition even though non-trivial we did not find useful in these examples. The issues are that for simple representations that we consider here this is automatically satisfied (for example $\Delta_{\ydiagram{1}}={(N^2-1)\over 2N(k+N)}$, $\Delta_{\ydiagram{1,1}}=\frac{(N-2) (N+1)}{N (k+N)}$, $\Delta_{\textbf{Adj}}={N\over N+k}\leq 1$). Therefore, this condition could have a chance to be useful for higher index symmetric and antisymmetric representations and exotic ones. However, most such examples constructed are for $T=0$ \cite{Kumar_2011}, but those theories tend to have a very large level $k$ since $a,b$ are scalars. Therefore, we would expect this to be more useful if a full 6d supergravity classification is considered and more exotic representation are considered for large $T$.

\section{Future directions}
In this work we have argued, using a combination of anomaly conditions and unitarity on BPS strings, that at least all the proposed 6d supergravity theories have an upper bound on the number of massless modes. Furthermore, completeness of spectrum led us to a new Swampland constraint that helps restrict the types of representations that can appear in a consistent theory of gravity. Those constraints helped us exclude theories that have no string theory realization and hence strengthening the validity of the SLP.  

  It would be interesting to generalize the finiteness argument for representation types as well.  In particular for abelian theories, the charged fields, even though finite in number, are known to have an infinite family of allowed charges \cite{Taylor_2018,raghuram2020automatic}.  We expect these, with the exception of a finite number of them,  to belong to the Swampland.   Therefore, it is  important to develop further techniques to rule these out.
  
  Finally, another interesting direction would be to try and enumerate all the gauge groups that actually do appear in the string landscape and provide an explanation of their appearance using only Swampland principles without specifying the particular UV completion.
  
  \section{Acknowledgments}
  We would like to thank Washington Taylor, Sheldon Katz, Hee-Cheol Kim, Guglielmo Lockhart and Noam D. Elkies,   for valuable discussions. The research of HCT and CV is supported in part by the NSF grant NSF PHY-2013858 and by a grant from the Simons Foundation (602883, CV).

\appendix
\section{Appendix}
 \subsection{Infinite Families}
 \label{app:inf}
 \ytableausetup{boxsize=0.7 em,aligntableaux = center}
The objective of this paper is to show that there are only finitely many massless modes for a 6d ${\cal N }=1$ theory and hence it is important to understand the possible infinite families  that could occur. As was discussed in Section \ref{sec:finite Landscape}, the only theories with one simple gauge factor and unbounded size are presented in the first two rows of Table \ref{table:1}.  However, they were both shown to be finite because of the existence of non-unitary BPS strings for arbitrarily large size. Therefore, of particular interest will be theories with multiple simple gauge   factors drawn from Table \ref{table:1} for which one can  reduce $H-V$  by gauging matter.

Specifically, we will argue that the only matter that can be gauged is fundamental matter. This fact was shown in \cite{Kumar_2010} to  be true for  $T<9$ and can be generalized for any $T$. By considering the group theory coefficients $A_R$ presented in  \cite{Kumar_2009} one can note that all representations except the fundamental contribute to $b_i\cdot b_j $ at least linear in  $N$. However, as was discussed extensively in Section \ref{sec:finite Landscape} $b_i$ vectors belong to the string charge lattice and consequently  are independent of $N$. Alternatively, one can also note from Table \ref{table:1} that this would not be possible for the specific theories.  For example, consider  the gauge group $G_N\times G_M$ with gauge factors picked from Table \ref{table:1} with  matter of the form  $(R_N,R_M)$ charged under both gauge groups in representations $R_N,R_M$ respectively. Assume now that at least one of representations is not fundamental and hence the only choices are symmetric /antisymmetric/adjoint. We may assume that $R_M$ is such a representation and hence $(R_N,R_M)=(R_N,{M(M\pm1)\over 2}\text{ or } \textbf{Adj})$. In order to form such a representation $SU(M)$ needs to have at least $\dim(R_N)$ representations in $M(M\pm1)/2\text{ or } \textbf{Adj}$  but that is not possible since there is only a finite number of those for each theory. Therefore,  we conclude that any matter charged under more than one gauge groups necessarily includes gauging of fundamental matter.

Furthermore, one should note from Table \ref{table:1}  that only matter charged under at most two groups can appear. For example, let us consider  a theory of the form $SU(N)\times SU(M)\times SU(K)$, 
even though trifundamental matter is possible for finite $N,M,K$ as shown in  \cite{Kumar_2010b}, it is not possible to construct it for unbounded size. This is because there needs to exist sufficiently many bifundamentals between
every pair of simple factors and the theories we are considering for arbitrary size do not. Alternatively, one could note that for example a trifundamental of $SU(N)\times SU(M)\times SU(K)$ would require at least $K$ bifundamentals $(N,M)$ but in this case $b_i\cdot b_j$
would also grow with $K$ which as discussed earlier is not consistent.

We also note that   $SU(N)/Sp(N)$ with $16\ \ydiagram{ 1}+ 2/1\ \ydiagram{ 1,1}$ or any other group of finite size can not be a factor because gauging bifundamental matter between unbounded size gauge groups requires also an unbounded number of hypermultiplets in the fundamental representation.
Therefore, we proceed by considering the possible ways three gauge groups from Table \ref{table:1} can combine and produce a theory with unbounded size not restricted by anomaly cancellation. 

\begin{itemize}
	\item Firstly, consider matter charged as $(N,M,1)+(1,M,K)+(N,1,K)$.

This type of  matter is only possible if the inequalities  $F_N\geq M+K,F_M\geq N+K, F_K\geq N+M\ $ are satisfied, where $F_N,F_M,F_K$ represent the number of hypermultiplets in the fundamental representation.
	From Table \ref{table:1} we know that there are three choices of each $F_N,F_M,F_K$ e.g.
	$F_N=N-8,N+8,2N$.
	Starting from the first possibility the above inequalities become:
	\begin{eqnarray}
	F_N=	N-8\geq M+K
	\end{eqnarray}
	and
	\begin{eqnarray}
	M+2K+8\leq N+K\leq F_M=2M 
	\end{eqnarray}
	with $F_M=2M $ being the only consistent choice for $K>0$.
	The third inequality becomes:
	\begin{eqnarray}
	2M+K+8\leq 	N+M\leq F_K=2K
	\end{eqnarray}
	again with $F_K=2K $ being the only consistent choice.

	Note that these inequalities can not be satisfied simultaneously for any combination of $N,M,K$ for large values 
	because the combination of the last two implies:
	\begin{eqnarray}
	N\leq {K+M\over 2 }
	\end{eqnarray}
	While the first 
	\begin{eqnarray}
	N\geq M+K+8
	\end{eqnarray}

One could also try  $F_N=N+8$ which leads to $F_K=2K, F_M=2M$ for large $N,M,K$ and hence one would need to satisfy $	N\leq {K+M\over 2 },	N\geq M+K-8$  for unbounded $N,M,K$  which is impossible. 
	
	Lastly, one can choose  $F_N=2N$ which forces $F_K=2K, F_M=2M$ and substituting this in the above inequalities imply that  $N=M=K  $. In particular, this theory has gravitational anomaly $H-V=3 N^2-3(N^2-1)=3$ and hence $T\leq 9$.	 
	More generally, we could consider such a loop  for arbitrary number of factors $SU(N)^k$  which has $H-V=k N^2-kN^2+k=k$ and hence $k\leq 12 , T\leq 9 $.

	\item We now move on to  charged matter of the form: $(N,M,1)+(1,M,K)$
	
	Lets us consider the three simple gauge factors $G_N\times G_M\times G_K$ with each component drawn from Table \ref{table:1}.
	Then assuming that no matter is charged under more than one gauge group we have that the leading contribution to $H-V$ is given by $c_1N^2+c_2M^2+c_3K^2$  with $c_i={1\over 2}\text{ or }1$ depending on the matter. Therefore, in order to make $H-V$ finite we can gauge matter in bifundamental representations which may eliminate the quadratic leading behavior of the contribution to the gravitational anomaly. As we did before the number of fundamental hypermultiplets  $F_N,F_M,F_K$ needs to satisfy  $F_N\geq M,F_M\geq N+ K, F_K\geq M $ with the leading behaviour $F_N\sim 2c_1 N,F_M\sim 2c_2 M,F_K\sim 2c_3 K$ and hence $2c_1 N \pm2(1-c_1)8\geq M, \quad 2c_2 M\pm2(1-c_2)8\geq N+ K, \quad 2c_3 K \pm2(1-c_3)8 \geq M $. Therefore, for large values of $M,N,K$ we have the leading terms $M\sim 2c_1 \alpha N$, \quad $K\sim 4c_1c_2  \alpha \beta N$ with $\alpha\leq 1,\beta <1$ and such that they satisfy the above inequalities.  This implies that the contribution for each bifundamental is $2c_1 \alpha N^2 + 8c_1^2c_2 \alpha^2 \beta N^2$ and hence this is the amount subtracted from $H-V$ when gauging $F_N,F_M,F_K$.  We thus  require that $c_1N^2+c_2M^2+c_3K^2=c_1N^2+c_2(2 c_1 a)^2N^2+16 c_3(c_1c_2 \alpha \beta )^2N^2=2c_1 \alpha N^2 + 8\alpha ^2 c_1^2c_2 \beta N^2$ in order to cancel the leading $N$ behavior of the gravitational anomaly. The equation  $N^2+4 c_2 c_1\alpha^2 N^2+16 c_3c_1c_2^2\alpha^2\beta^2N^2=2 \alpha N^2 + 8\alpha^2c_1c_2 \beta N^2$ has the following solutions that eliminate the quadratic behavior from $H-V$: $(\alpha,\beta,c_1,c_2,c_3)=(1,{1\over 2},{1\over 2},1,{1\over 2}),(1,{1\over 2},1,1,{1\over 2})$.

	Starting from the first solutions and enforcing  $F_N\geq M,F_M\geq N+K, F_K\geq M\ $ we can find the consistent solutions. 
	The possibilities as indicated from $(\alpha,\beta,c_1,c_2,c_3)=(1,{1\over 2},{1\over 2},1,{1\over 2})$ are :
	$F_N=N-8,N+8$, \quad $F_M=2M=2N$, \quad $F_K=N-8, \quad N+8$.
	The first case requires $F_N=N-8\geq M$, $M+K+8\leq N+K\leq 2M $ and $K+8\leq 	M\leq F_K$ we have $F_K=K+8$.

	The second choice of $F_N$ is similar to above with $N+8\geq M$,$M+K-8\leq N+K\leq 2M$ and $K-8\leq 	M\leq F_K$. Consistency then requires $F_K=K-8$ (which is identical to the previous case)or $F_K=K+8$.
	
	The second category with $(\alpha,\beta,c_1,c_2,c_3)=(1,{1\over 2},1,1,{1\over 2})$  requires $2N\geq M$, $M\leq 2K$, $M\leq N+K\leq F_M$ with $F_M=2M$.
	
	All the theories that satisfy those  conditions are summarized in Table 		\ref{table:N}. The last column indicates the contribution to $H-V$ of each theory, one can see that although the quadratic terms are  eliminated there can still be linear terms.
	This could have also been deduced by looking at each individual contribution of $H-V$ from Table \ref{table:1} and noticing that the theories with $c_i=1/2$ need to come in pairs such that the linear contributions cancel because gauging bifundamental matter does not affect the linear terms as long as the size  is not restricted to be even. 
	Therefore,  only four theories do not have their dimensions restricted by anomalies.

	\begin{table}[h!]
		\scalebox{0.8}{
			\begin{tabular}{|l|l|l|}
				\hline
				$SU(N+8)\times SU(N)\times SU(N-8)$ & (\ydiagram{1},${\ydiagram{1}}$,1) +(1,\ydiagram{1},${\ydiagram{1}}$)+(\ydiagram{2},1,1)& -53 \\ &  +(1,1,\ydiagram{1,1}) & \\\hline

				$SU(N-8)\times SU(N)\times SU(N-8)$ & (\ydiagram{1},${\ydiagram{1}}$,1) +(1,\ydiagram{1},${\ydiagram{1}}$)+(\ydiagram{1,1},1,1)& $15 N-53$\\ & +(1,1,\ydiagram{1,1})+16(1,\ydiagram{1},1)& \\ \hline

				$SU(N-8)\times SU(N)\times SU(N+8)$ & (\ydiagram{1},${\ydiagram{1}}$,1) +(1,\ydiagram{1},${\ydiagram{1}}$)+(\ydiagram{1,1},1,1)& $  $\\ & +(1,1,\ydiagram{1,1})+16(1,1,\ydiagram{1})& $15 N+67$ \\ \hline

				$SU(N)\times SU(2N)\times SU(N)$ & (\ydiagram{1},${\ydiagram{1}}$,1) +(1,\ydiagram{1},${\ydiagram{1}}$)& $15 N-3$\\ & +$8$ (1,${\ydiagram{1}}$,1)+(1,${\ydiagram{1,1}}$,1)&\\ \hline

				$SU(N)\times SU(2N)\times SU(N+8)$ & $(\ydiagram{ 1},\ydiagram{1},1)$             +$(1,\ydiagram{ 1},\ydiagram{1})$                                         & $15 N+61$  \\ 
				& +(1,${\ydiagram{1,1}}$,1) +$16$$(1,1,\ydiagram{1})$&	\\ \hline

				$SO(M+8)\times SU(M)\times SU(M-8)$ & $(\ydiagram{ 1},\ydiagram{1},1)$             +$(1,\ydiagram{ 1},\ydiagram{1})$                                         & $-58$  \\ 
				& +(1,1,${\ydiagram{1,1}}$) &	\\ \hline

				$SU(M+8)\times SU(M)\times Sp((M-8)/2)$ & $(\ydiagram{ 1},\ydiagram{1},1)$             +$(1,\ydiagram{ 1},\ydiagram{1})$                                         & $-58$  \\ 
				& +(${\ydiagram{2}}$,1,1) &	\\ \hline

				$SO(M+8)\times SU(M)\times Sp((M-8)/2)$ & $(\ydiagram{ 1},\ydiagram{1},1)$             +$(1,\ydiagram{ 1},\ydiagram{1})$                                         & $-57$ 	\\ \hline

				$Sp((M-8)/2)\times SU(M)\times Sp((M-8)/2)$ & $(\ydiagram{ 1},\ydiagram{1},1)$             +$(1,\ydiagram{ 1},\ydiagram{1})$      +16 $(1,\ydiagram{1},1)$                                   & $15 N-57$  	\\ \hline

				$SU(N)\times Sp(M)\times SU(N)$ & $(\ydiagram{ 1},\ydiagram{1},1)$             +$(1,\ydiagram{ 1},\ydiagram{1})$      +8 $(1,\ydiagram{1},1)$                                   & $15 N-2$  	\\ \hline
						\end{tabular}}
		\caption{Theories with three simple gauge factors and $H-V$ at most linear in $N$.}
		\label{table:N}
	\end{table}

	In a similar way we can extend this analysis to more than 3 gauge factors. We may start by considering a linear chain of gauge groups of the form  $G_1\times G_2\times\cdots \times G_k$ with bifundamental matter charged under every adjacent pair of groups $G_i$ each of size $N_i=a_i N+c_i$. Let us start by considering $G_i$ with  matter in $F_{N_i}=N_i\pm 8$. Then  the size  of each adjacent gauge group is bounded as: $F_{N_i}\geq N_{i+1}+N_{i-1}$ which implies that $a_iN +c_i\pm 8 \geq a_{i-1}N+a_{i+1}N+c_{i+1}+c_{i-1} $.
	For large $N$ this inequality can be translated to keeping only the  linear terms in $N$ given  by:
	\begin{eqnarray}
	a_iN \geq a_{i-1}N+a_{i+1}N
	\end{eqnarray}
	Assuming the same type of matter for the  adjacent groups we have 
	\begin{eqnarray}
	a_{i+1}N \geq a_{i}N+a_{i+2}N, \ a_{i-1}N \geq a_{i}N+a_{i-2}N
	\end{eqnarray}
	
	Adding the first inequality and the last two gives:
	\begin{eqnarray}
	0\geq a_{i} +a_{i+2}+a_{i-2} 
	\end{eqnarray}
	which can not be satisfied for positive $a_i$'s.
	We can instead consider a different type of matter for one of the gauge groups($F_{i+1}=2N_{i+1}$) satisfying:
	\begin{eqnarray}
	2a_{i+1}N \geq a_{i}N+a_{i+2}N, \ a_{i-1}N \geq a_{i}N+a_{i-2}N
	\end{eqnarray}
	Combining these equations gives:
	\begin{eqnarray}
	0\geq{1\over 2 }a_{i}+{1\over 2 }a_{i+2}+a_{i-2}
	\end{eqnarray}
	which is also not satisfied for positive $a_i$'s.
	
	Next we can consider also  $F_{i-1}=2N_{i-1}$ which needs the following inequalities to be satisfied:
	\begin{eqnarray}
	2a_{i+1}N \geq a_{i}N+a_{i+2}N,	2a_{i-1}N \geq a_{i}N+a_{i-2}N
	\end{eqnarray}
Which can combine to:
	\begin{eqnarray}\label{eq:212}
	0\geq{1\over 2 }a_{i+2} +{1\over 2 }a_{i-2}
	\end{eqnarray}
	
	Similarly, if instead one had: $F_{i-1}=N_{i-1},F_{i+1}=N_{i+1}, F_{i}=2N_i$
	then  the following inequalities need to be satisfied:
		\begin{eqnarray}
		\hskip -1cm
2	a_iN \geq a_{i-1}N+a_{i+1}N,\ 	a_{i+1}N \geq a_{i}N+a_{i+2}N, \ a_{i-1}N \geq a_{i}N+a_{i-2}N
	\end{eqnarray}
And hence
		\begin{eqnarray}\label{eq:121}
	0\geq a_{i+2} + a_{i-2}
	\end{eqnarray}
	as above, which can not be satisfied for theories with more than three gauge groups. However, if the chain has only three gauge groups $a_{i+2},a_{i-2}=0$ and hence the \ref{eq:121}, \ref{eq:212} inequalities are satisfied. In particular, both  theories were found earlier in Table \ref{table:N} but the latter had $H-V$ linear in $N$ while the former had $H-V$ constant as desired. Therefore, any infinite family should only have gauge groups away from the edges of type $SU(N)+2N\ydiagram{1}$. However, even though the other simple gauge factors do not appear away from the edges of the chain, there is nothing wrong with them being the first and last  factors. For example, if $G_i$ was the first gauge group then $a_{i-1}=a_{i-2}=0$ and hence the inequalities become: $a_i\geq a_{i+1} , 2a_{i+1} \geq a_{i}+a_{i+2}$ which can be solved.
	
	Lastly, let us look at theories with more than three gauge groups starting with $SU(N)+2N\ydiagram{1}$.
	In this case we require the following inequalities to be satisfied for the $SU(N)\times SU(a_1N+c_1)\times \cdots SU(a_k N+c_k)$ gauge group: 
	\begin{eqnarray}\hskip -1cm
	2  \geq a_{1},2a_1 \geq 1+a_{2},\ 2a_{2} \geq a_{1}+a_{3},\cdots , 2a_{k-1}\geq a_k+a_{k-2}, F_K \geq a_{k-1}N
	\end{eqnarray}

	We can start by investigating the different possible solutions for these inequalities.  
	Let $a_1=1$ then  there is a unique choice for all $a_i=1$ but there are  $N$ fundamental hypermultiplets for the first and last gauge groups which have not been gauged and consequently  give quadratic contributions to $H-V$. 
	However, if   $a_1=2$ then $a_2=2,3$, for $a_2=2$ then all $a_i=2$ but if $a_2=3$ then trying to saturate all inequalities we get increasing $a_i$: $a_0=1,a_1=2,a_2=3,a_3=4,\cdots a_{k-1}=k, a_k=k+1$. All these theories have a large number $\sim N$ of ungauged matter in the fundamental representation which lead to quadratic diverge in $H-V$ for large $N$ irrespectively of the choice of $F_K$. 
	Therefore, the only theories that may give potential infinite families are those with gauge groups of the form:
	\begin{eqnarray}
	G_1\times SU(N_2)\times \cdots \times SU(N_{k-1})\times G_k
	\end{eqnarray}
	with $G_k$ any group with matter that contains  $(N_i\pm 8)\ydiagram{1}$. All these cases are studied extensively in Section \ref{sec:finite Landscape}.
	
Here we have considered only linear chains of groups that could be potential infinite families.

\end{itemize}

\subsection{Non-linear chains}\label{appnon}
Consider a theory that has at least one of the groups connected to $n$ other groups. 
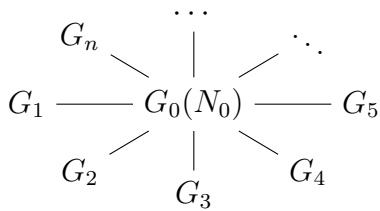
\newline
{\begin{tikzpicture}
	\node (v1) at (0.3,0) {$G_1$};
	\node (v5) at (4.7,0) {$G_5$};
	\node (v7) at (4,-0.9) {$G_4$};
	\node (v9) at (2.5,-1.2) {$G_3$};
	\node (v4) at (2.5,1.2) {$\cdots$};
	\node (v2) at (2.5,0) {$G_0(N_0)$};
	\node(v3) at (4,0.9) {$\ddots$};
	\node(v6) at (1,-0.9) {$G_2$};
	\node(v8) at (1,0.9) {$G_n$};
	\draw  (v1) edge (v2);
	\draw  (v3) edge (v2);
	\draw  (v4) edge (v2);
	\draw  (v5) edge (v2);
	\draw  (v6) edge (v2);
	\draw  (v7) edge (v2);
	\draw  (v8) edge (v2);
	\draw  (v9) edge (v2);
	\end{tikzpicture}} 
\newline
We know from Table \ref{table:1} each group $G_i(N_i)$ depends on a parameter $N_i$ that control its size and the matter associated to that group can be either $N_i\pm 8$ or $2N_i$ fundamental hypermultiplets, which we will label  as $k_i N_i\pm (2-k_i)8$ with $k_i=1\text{ or } 2$. 
Therefore, assuming that there is matter charged under each adjacent group  the following inequalities for the leading contributions should hold:
\begin{eqnarray}
\sum_{i=1}^k N_i\leq F_{N_0}=k_0 N_0 ,  N_0 \leq F_{N_i}=k_i N_i
\end{eqnarray}
We can  combine them to get 
\begin{eqnarray}
\sum_i{1\over k_i}\leq k_0
\end{eqnarray}
The largest number of gauge groups $i$ is reached when $k_i=2$ and  $k_0 =2$ for which  $i=4$. Therefore, $n=4$ is the largest number of adjacent groups one can have. We have already studied the cases with $n=0,1,2$ which corresponds to the linear chains. For $n\geq3$ one would require $k_0=2$ and hence the only possibilities are $(k_1,k_2,k_3)=(1,2,2),(2,2,2)$ or $(k_1,k_2,k_3,k_3)=(2,2,2,2)$. 

For $n=4$:
\begin{eqnarray}
N_1+N_2+N_3+N_4=2N_0
\end{eqnarray}
and $N_i\geq N_0/2$ then the unique solution is:
\newline
{\begin{tikzpicture}
	\node (v1) at (0,0.5) {$SU(N)$};
	\node (v4) at (5,0.5) {$SU(N)$};
	\begin{scope}[shift={(1.5,0)}]
	\node (v2) at (1,0.5) {$SU(2N)$};
	\node(v3) at (1,1.5) {$SU(N)$};
	\node(v6) at (1,-0.5) {$SU(N)$};
	\end{scope}
	\draw  (v1) edge (v2);
	\draw  (v3) edge (v2);
	\draw  (v4) edge (v2);
	\draw  (v6) edge (v2);
	\end{tikzpicture}} 
\newline
with $H-V=5 $ and hence $T\leq 9 $.

For $n=3$:
\begin{eqnarray}
{3\over 2 }N_0\leq N_1+N_2+N_3=2N_0
\end{eqnarray}

and $N_i\geq N_0/2$ from which one could construct the following infinite families of theories :

{\begin{tikzpicture}
	\node (v11) at (0,0.5){};
	\begin{scope}[shift={(0,0)}]
	\node (v1) at (0,0.5) {$SU(2N)$};
	\node (v8) at (0,01.5) {$SU(N)$};
	\node (v9) at (-2.5,0.5) {$SU(N)$};
	\begin{scope}[shift={(3,0)}]
	\node (v4) at (5,0.5) {$SU(N)$};
	\node (v2) at (2.5,0.5) {$SU(2N)$};
	\node(v3) at (2.5,1.5) {$SU(N)$};
	\end{scope}
	\node(v7) at (2.7,0.5) {$SU(2N)^m$};
	\draw[dashed]   (v1) edge (v7);
	\draw[dashed]  (v2) edge (v7);
	\draw  (v3) edge (v2);
	\draw  (v4) edge (v2);
	\draw  (v9) edge (v1);
	\draw  (v8) edge (v1);
	\end{scope}
	\end{tikzpicture}} 
\newline
with  $H-V=6+m$ which implies that $T\leq9$. 
\newline

{\begin{tikzpicture}
	\node (v11) at (0,0.5){};
	\begin{scope}[shift={(2,0)}]
	\node (v1) at (0,0.5) {$SU(2N)$};
		\node (v5) at (-2.5,0.5) {$SU(N)$};
	\node (v4) at (5,0.5) {$SU(2N)$};
	\node (v6) at (7.5,0.5) {$SU(N)$};
	\node (v2) at (2.5,0.5) {$SU(3N)$};
	\node(v3) at (2.5,1.5) {$SU(2N)$};
	\node(v7) at (2.5,2.5) {$SU(N)$};
	\draw  (v1) edge (v2);
		\draw  (v1) edge (v5);
	\draw  (v3) edge (v2);
	\draw  (v3) edge (v7);
	\draw  (v4) edge (v2);
	\draw  (v4) edge (v6);
		\end{scope}
	\end{tikzpicture}} 
\newline
with $H-V=7$ which implies that $T\leq 9$.
\newline

{\begin{tikzpicture}
	\node (v9) at (0,0) {};
	\begin{scope}[shift={(-2,0)}]
	\node (v1) at (2,-0.5) {$SU(2N)$};
	\node (v4) at (4,0.5) {$SU(3N)$};
	\node (v6) at (6,0.5) {$SU(2N)$};
	\node (v7) at (8,0.5) {$SU(N)$};
	\node (v9) at (-4,0.5) {$SU(N)$};
	\node (v2) at (2,0.5) {$SU(4N)$};
	\node(v3) at (0,0.5) {$SU(3N)$};
	\node(v5) at (-2,0.5) {$SU(2N)$};
	\end{scope}
	\draw  (v1) edge (v2);
	\draw  (v3) edge (v2);
	\draw  (v4) edge (v2);
	\draw  (v5) edge (v3);
	\draw  (v6) edge (v4);
	\draw  (v6) edge (v7);
	\draw  (v5) edge (v9);
	\end{tikzpicture}} 
\newline
with $H-V=8$ which implies that $T\leq9$.
\newline

{\begin{tikzpicture}
	\node (v9) at (0,0) {};
	\begin{scope}[shift={(-2,0)}]
	\node (v1) at (2,-0.5) {$SU(3N)$};
	\node (v4) at (4,0.5) {$SU(5N)$};
	\node (v6) at (6,0.5) {$SU(4N)$};
	\node (v7) at (8,0.5) {$SU(3N)$};
	\node (v8) at (10,0.5) {$SU(2N)$};
	\node (v9) at (12,0.5) {$SU(N)$};
	\node (v2) at (2,0.5) {$SU(6N)$};
	\node(v3) at (0,0.5) {$SU(4N)$};
	\node(v5) at (-2,0.5) {$SU(2N)$};
	\end{scope}
	\draw  (v1) edge (v2);
	\draw  (v3) edge (v2);
	\draw  (v4) edge (v2);
	\draw  (v5) edge (v3);
	\draw  (v6) edge (v4);
	\draw  (v6) edge (v7);
	\draw  (v7) edge (v8);
	\draw  (v8) edge (v9);
	\end{tikzpicture}} 
\newline
with $H-V=9$ which implies that $T\leq9$.
\newline
All these theories have $a^2=9-T$, $b_i^2=-2$,$b_i\cdot b_j =1\text{ or 0}$ for $i \neq j $ depending on the inner product pattern presented on the diagrams above. 
These theories together with the cyclic $SU(N)$ we found earlier \ref{app:inf} have anomaly lattices equal to the negative of the extended Cartan matrices from affine ADE.

The other case with 
\begin{eqnarray}
N_1+N_2+N_3=2N_0
\end{eqnarray}
and $N_1\geq N_0,N_{i\neq 1}\geq N_0/2$ does not seem to give infinite families.

We note that the list of theories might not be exhaustive since we have not considered any more exotic configurations where for example loops could appear.

\bibliographystyle{JHEP}
\bibliography{6dSugra}

\providecommand{\href}[2]{#2}\begingroup\raggedright\begin{thebibliography}{10}

\bibitem{Vafa:2005ui}
C.~Vafa, {\it {The String landscape and the swampland}},
  \href{http://arxiv.org/abs/hep-th/0509212}{{\tt hep-th/0509212}}.

\bibitem{Yau:1991}
S.-T. Yau, {\it {A Review of {C}omplex {D}ifferential {G}eometry}},  {\em Proc.
  of Symp. in Pure Math.} {\bf 52} (1991) 619--625.

\bibitem{Adams:2010zy}
A.~Adams, O.~DeWolfe, and W.~Taylor, {\it {String universality in ten
  dimensions}},  {\em Phys. Rev. Lett.} {\bf 105} (2010) 071601,
  [\href{http://arxiv.org/abs/1006.1352}{{\tt arXiv:1006.1352}}].

\bibitem{Green:1984sg}
M.~B. Green and J.~H. Schwarz, {\it {Anomaly {C}ancellation in {S}upersymmetric
  {$D=$10} {G}auge Theory and Superstring Theory}},  {\em Phys. Lett.} {\bf
  149B} (1984) 117--122.

\bibitem{Kim:2019aa}
H.-C. Kim, G.~Shiu, and C.~Vafa, {\it Branes and the swampland},  {\em Phys.
  Rev. D} {\bf 100} (2019) 066006, [\href{http://arxiv.org/abs/1905.08261}{{\tt
  arXiv:1905.08261}}].

\bibitem{Kim_2020d}
H.-C. Kim, H.-C. Tarazi, and C.~Vafa, {\it Four-dimensional $\mathcal{N}=$4
  {SYM }theory and the swampland},  {\em Physical Review D} {\bf 102} (Jul,
  2020).

\bibitem{Kumar_2010}
V.~Kumar, D.~R. Morrison, and W.~Taylor, {\it Global aspects of the space of
  6{D} $ \mathcal{N} = 1 $ supergravities},  {\em Journal of High Energy
  Physics} {\bf 2010} (Nov, 2010).

\bibitem{Taylor_2019}
W.~Taylor and A.~P. Turner, {\it Generic matter representations in 6{D}
  supergravity theories},  {\em Journal of High Energy Physics} {\bf 2019}
  (May, 2019).

\bibitem{Kumar_2009}
V.~Kumar and W.~Taylor, {\it A bound on 6d $\mathcal{N}$=1 supergravities},
  {\em Journal of High Energy Physics} {\bf 2009} (Dec, 2009) 050–050.

\bibitem{Kumar_2011}
V.~Kumar, D.~S. Park, and W.~Taylor, {\it {6D} supergravity without tensor
  multiplets},  {\em Journal of High Energy Physics} {\bf 2011} (Apr, 2011).

\bibitem{Morrison_2012}
D.~Morrison and W.~Taylor, {\it Classifying bases for 6{D} {F}-theory models},
  {\em Open Physics} {\bf 10} (Jan, 2012).

\bibitem{Morrison_2012b}
D.~Morrison and W.~Taylor, {\it Toric bases for 6{D} {F}-theory models},  {\em
  Fortschritte der Physik} {\bf 60} (May, 2012) 1187–1216.

\bibitem{Kumar_2010b}
V.~Kumar, D.~R. Morrison, and W.~Taylor, {\it Mapping 6{D} $ \mathcal{N} = 1 $
  supergravities to {F}-theory},  {\em Journal of High Energy Physics} {\bf
  2010} (Feb, 2010).

\bibitem{raghuram2020automatic}
N.~Raghuram, W.~Taylor, and A.~P. Turner, {\it Automatic {E}nhancement in 6{D}
  {S}upergravity and {F}-theory models},  2020.

\bibitem{Lee_2019}
S.-J. Lee and T.~Weigand, {\it {Swampland bounds on the {A}belian gauge
  sector}},  {\em Physical Review D} {\bf 100} (Jul, 2019).

\bibitem{Park_2012}
D.~S. Park and W.~Taylor, {\it {Constraints on 6{D} supergravity theories with
  abelian gauge symmetry}},  {\em Journal of High Energy Physics} {\bf 2012}
  (Jan, 2012).

\bibitem{Cveti__2020}
M.~Cvetič, M.~Dierigl, L.~Lin, and H.~Y. Zhang, {\it String {U}niversality and
  {N}on-{S}imply-{C}onnected {G}auge {G}roups in 8{D}},  {\em Physical Review
  Letters} {\bf 125} (Nov, 2020).

\bibitem{Dierigl_2021}
M.~Dierigl and J.~J. Heckman, {\it Swampland cobordism conjecture and
  non-{A}belian duality groups},  {\em Physical Review D} {\bf 103} (Mar,
  2021).

\bibitem{AlvarezGaume:1983ig}
L.~Alvarez-Gaume and E.~Witten, {\it {Gravitational Anomalies}},  {\em Nucl.
  Phys. B} {\bf 234} (1984) 269.

\bibitem{Montero_2021}
M.~Montero and C.~Vafa, {\it Cobordism conjecture, anomalies, and the {S}tring
  {L}amppost principle},  {\em Journal of High Energy Physics} {\bf 2021} (Jan,
  2021).

\bibitem{Garc_a_Etxebarria_2017}
I.~García-Etxebarria, H.~Hayashi, K.~Ohmori, Y.~Tachikawa, and K.~Yonekura,
  {\it 8d gauge anomalies and the topological {G}reen-{S}chwarz mechanism},
  {\em Journal of High Energy Physics} {\bf 2017} (Nov, 2017).

\bibitem{hamada20218d}
Y.~Hamada and C.~Vafa, {\it 8d supergravity, {R}econstruction of {I}nternal
  {G}eometry and the {S}wampland},  2021.

\bibitem{font2021exploring}
A.~Font, B.~Fraiman, M.~Graña, C.~A. Núñez, and H.~P.~D. Freitas, {\it
  Exploring the landscape of {CHL} strings on $t^d$},  2021.

\bibitem{Font_2020}
A.~Font, B.~Fraiman, M.~Graña, C.~A. Núñez, and H.~P. De~Freitas, {\it
  Exploring the landscape of heterotic strings on $t^d$},  {\em Journal of High
  Energy Physics} {\bf 2020} (Oct, 2020).

\bibitem{Klevers_2017}
D.~Klevers, D.~R. Morrison, N.~Raghuram, and W.~Taylor, {\it Exotic matter on
  singular divisors in {F}-theory},  {\em Journal of High Energy Physics} {\bf
  2017} (Nov, 2017).

\bibitem{Morrison_2012f}
D.~R. Morrison and W.~Taylor, {\it Matter and singularities},  {\em Journal of
  High Energy Physics} {\bf 2012} (Jan, 2012).

\bibitem{Sagnotti:1992qw}
A.~Sagnotti, {\it {A Note on the {G}reen-{S}chwarz mechanism in open string
  theories}},  {\em Phys. Lett.} {\bf B294} (1992) 196--203,
  [\href{http://arxiv.org/abs/hep-th/9210127}{{\tt hep-th/9210127}}].

\bibitem{Seiberg_2011}
N.~Seiberg and W.~Taylor, {\it Charge lattices and consistency of 6{D}
  supergravity},  {\em Journal of High Energy Physics} {\bf 2011} (Jun, 2011).

\bibitem{Monnier_2019}
S.~Monnier and G.~W. Moore, {\it Remarks on the {G}reen–{S}chwarz terms of
  six-dimensional supergravity theories},  {\em Communications in Mathematical
  Physics} {\bf 372} (Feb, 2019) 963–1025.

\bibitem{Cheung_2017}
C.~Cheung and G.~N. Remmen, {\it Positivity of {C}urvature-{S}quared
  {C}orrections in {G}ravity},  {\em Physical Review Letters} {\bf 118} (Feb,
  2017).

\bibitem{Hamada_2019}
Y.~Hamada, T.~Noumi, and G.~Shiu, {\it {W}eak {G}ravity {C}onjecture from
  {U}nitarity and {C}ausality},  {\em Physical Review Letters} {\bf 123} (Jul,
  2019).

\bibitem{Banks:2010zn}
T.~Banks and N.~Seiberg, {\it {Symmetries and {S}trings in {F}ield {T}heory and
  {G}ravity}},  {\em Phys. Rev.} {\bf D83} (2011) 084019,
  [\href{http://arxiv.org/abs/1011.5120}{{\tt arXiv:1011.5120}}].

\bibitem{Polchinski:2003bq}
J.~Polchinski, {\it {Monopoles, duality, and string theory}},  {\em Int. J.
  Mod. Phys.} {\bf A19S1} (2004) 145--156,
  [\href{http://arxiv.org/abs/hep-th/0304042}{{\tt hep-th/0304042}}].
  [,145(2003)].

\bibitem{Schwarz_1996}
J.~H. Schwarz, {\it Anomaly-free supersymmetric models in six dimensions},
  {\em Physics Letters B} {\bf 371} (Mar, 1996) 223–230.

\bibitem{Polchinski_2004}
J.~Polchinski, {\it Monopoles, {D}uality, and {S}tring {T}heory},  {\em
  International Journal of Modern Physics A} {\bf 19} (Feb, 2004) 145–154.

\bibitem{Banks_2011}
T.~Banks and N.~Seiberg, {\it Symmetries and strings in field theory and
  gravity},  {\em Physical Review D} {\bf 83} (Apr, 2011).

\bibitem{harlow2019symmetries}
D.~Harlow and H.~Ooguri, {\it Symmetries in quantum field theory and quantum
  gravity},  2019.

\bibitem{Duff_1996}
M.~Duff, H.~Lü, and C.~Pope, {\it Heterotic phase transitions and
  singularities of the gauge dyonic string},  {\em Physics Letters B} {\bf 378}
  (Jun, 1996) 101–106.

\bibitem{kumar2009string}
V.~Kumar and W.~Taylor, {\it String universality in {S}ix {D}imensions},  2009.

\bibitem{martini20156d}
G.~Martini and W.~Taylor, {\it 6{D} {F}-theory models and elliptically fibered
  {C}alabi-{Y}au threefolds over semi-toric base surfaces},  2015.

\bibitem{Taylor_2012h}
W.~Taylor, {\it On the {H}odge structure of elliptically fibered {C}alabi-{Y}au
  threefolds},  {\em Journal of High Energy Physics} {\bf 2012} (Aug, 2012).

\bibitem{Heckman_2019}
J.~J. Heckman and T.~Rudelius, {\it Top down approach to 6{D} {SCFTs}},  {\em
  Journal of Physics A: Mathematical and Theoretical} {\bf 52} (Feb, 2019)
  093001.

\bibitem{Katz_2020}
S.~Katz, H.-C. Kim, H.-C. Tarazi, and C.~Vafa, {\it Swampland constraints on 5d
  $ \mathcal{N} $ = 1 supergravity},  {\em Journal of High Energy Physics} {\bf
  2020} (Jul, 2020).

\bibitem{DiFrancesco:1997nk}
P.~Di~Francesco, P.~Mathieu, and D.~Senechal, {\em {Conformal {F}ield
  {T}heory}}.
\newblock Graduate Texts in Contemporary Physics. Springer-Verlag, New York,
  1997.

\bibitem{Taylor_2018}
W.~Taylor and A.~P. Turner, {\it {An infinite swampland of U(1) charge spectra
  in { 6D} supergravity theories}},  {\em Journal of High Energy Physics} {\bf
  2018} (Jun, 2018).

\end{thebibliography}\endgroup

\end{document}